\documentclass[%
reprint,
 amsmath,amssymb,
 aps,
prstab,
floatfix,
]{revtex4-2}

\usepackage{graphicx}
\usepackage{dcolumn}
\usepackage{bm}
\usepackage{hyperref}
\usepackage{siunitx}
\newcommand{\E}{\mathcal{E}}
\newcommand{\F}{\mathcal{F}}

\begin{document}


\title{Longitudinal mode-coupling instabilities of proton bunches \\in the CERN Super Proton Synchrotron}

\author{Ivan Karpov}
\email{ivan.karpov@cern.ch}
\affiliation{CERN, CH-1211 Geneva, Switzerland}%





\date{\today}

\begin{abstract}

In this paper, we study single-bunch instabilities observed in the CERN Super Proton Synchrotron~(SPS). According to the linearized Vlasov theory, radial or azimuthal mode-coupling instabilities result from a coupling of bunch-oscillation modes, which belong to either the same or adjacent azimuthal modes, respectively. We show that both instability mechanisms exist in the SPS by applying the Oide-Yokoya approach to compute van Kampen modes for the realistic longitudinal impedance model of the SPS. The results agree with macroparticle simulations and are consistent with beam measurements. In particular, we see that the uncontrolled longitudinal emittance blow-up of single bunches observed before the recent impedance reduction campaign (2018-2021) is due to the radial mode-coupling instability. Unexpectedly, this instability is as strong as the azimuthal mode-coupling instability, which is possible in the SPS for other combinations of bunch length and intensity. We also demonstrate the significant role of rf nonlinearity and potential-well distortion in determining these instability thresholds. Finally, we discuss the effect of the recent impedance reduction campaign on beam stability in single- and double-rf configurations.
\end{abstract}

\maketitle


\section{Introduction}\label{sec:intro}

Longitudinal single-bunch instability is a possible performance limitation in many synchrotrons and its mechanism is a subject of various studies since long time \cite{Sacherer:1977, Besnier:1979, Wang1980, YHChin:1983, Krinsky1983, Laclare:1987, Garnier:1987, OY:1990, Oide:1995, Chao:1994, Ng:1995, Dyachkov:1995, Mosnier:1999, Cai2011, Lindberg2017, Blednykh2018}. The standard approach to evaluate beam stability is based on a solution of the linearized Vlasov equation for a small initial perturbation of a stationary distribution function. To simplify the analysis, the modification of a stationary potential well by self-induced fields, called potential-well distortion (PWD), is often neglected. The only possible mechanism of longitudinal single-bunch instability without PWD, a coupling of different azimuthal modes, was proposed by Sacherer~\cite{Sacherer:1977}. 

Another type of instability can be caused by asymmetry of the potential well due to PWD, resulting in a coupling of two radial modes within one azimuthal mode~\cite{OY:1990, Oide:1995}. An explicit condition required for this instability to occur was found for the double-waterbag model~\cite{Chao:1994}. For an  impedance model consisting of one broad-band resonator with frequency $f_r=\omega_r/2\pi$, the instability thresholds computed with and without PWD are similar for $\omega_r\sigma\gtrsim0.4$, where $\sigma$ is the rms bunch length,~\cite{OY:1990}. This result was also confirmed in calculations based on the orthogonal polynomial expansion~\cite{Cai2011}. 
The azimuthal mode-coupling was also found in the self-consistent analysis of electron bunches for $\omega_r\sigma\approx \pi$~\cite{Mosnier:1999}. 

Similar to electron bunches, the thresholds of the single-bunch instability for proton bunches are often computed neglecting bunch asymmetry due to PWD and rf nonlinearity, as for example, in~\cite{Ng:1995, MetralMigliorati:2020}, and thus only azimuthal mode-coupling instability was found. 
To our best knowledge, for proton bunches, so far a radial mode-coupling instability was not observed in measurements nor in calculations.

\begin{figure}[tb]
\begin{center}
\includegraphics{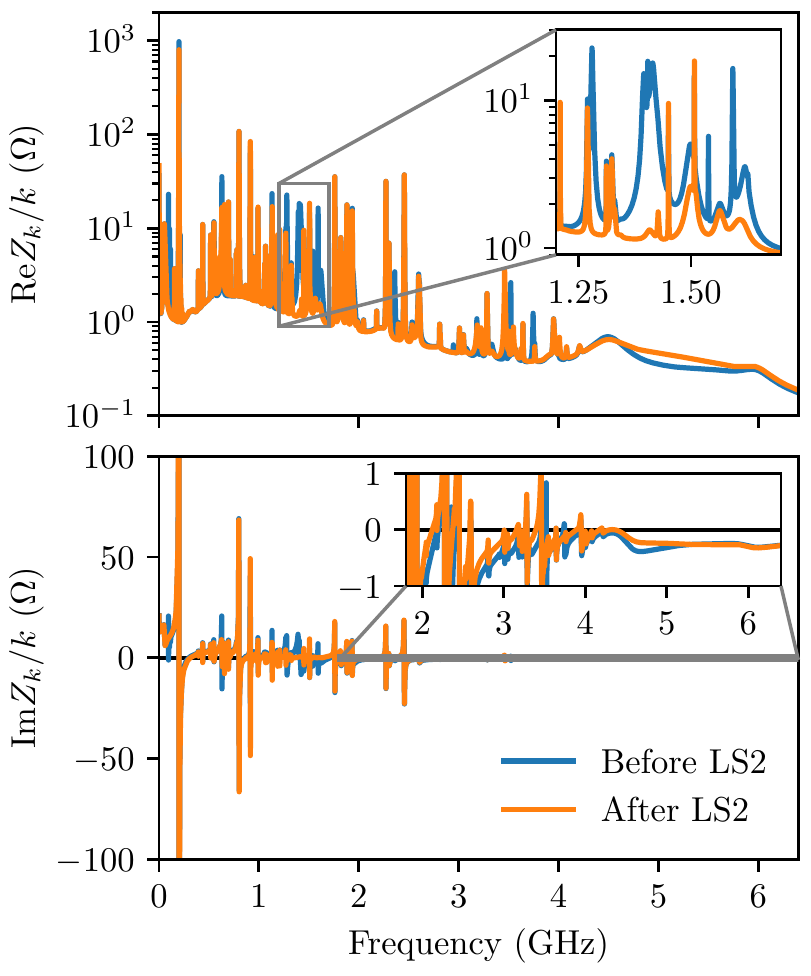}
\caption{SPS impedance model separated by real (top) and imaginary (bottom) parts before and after the impedance reduction campaign during the 2nd Long Shutdown (LS2), which ended in March 2021.}
\label{fig:sps_impedance}
\end{center}
\end{figure}

In the SPS, the longitudinal instability of single proton bunches occurs during the acceleration ramp. The attempts to cure this instability by reducing the voltage in a single rf system and thus increasing the synchrotron frequency spread for a constant longitudinal emittance were not successful. Instead, a higher rf voltage was more beneficial~\cite{Lasheen:2017}. In operation, this instability is cured by the application of a higher-harmonic (HH) rf system. 
Due to strong frequency dependence of the SPS impedance~\cite{SPS:2021}~(Fig.~\ref{fig:sps_impedance}), the observed instability was mainly studied in macroparticle simulations using the code {\footnotesize BLonD}~\cite{Blond:2021}. The latest results of simulations through the ramp are consistent with measurements and the agreement has been improved with the refined impedance model~\cite{Repond:2019}. 

In the present work, the mechanism of the SPS single-bunch instability is studied using code {\footnotesize MELODY}~\cite{Melody:2021} which is able to find in a fully self-consistent way the numerical solutions of the semianalytic matrix equations derived from the Vlasov equation for the full SPS impedance model. 
We show that the previously observed instability during the ramp was due to the coupling of multiple radial modes within one azimuthal mode. 
For a specific set of bunch parameters, we find an instability caused by the coupling of neighboring azimuthal modes for which Landau damping is lost. The main results are confirmed by macroparticle simulations with {\footnotesize BLonD} and are consistent with previous measurements~\cite{Lasheen:2017}.

The paper is organized as follows. In Sec.~\ref{sec: main}, we briefly discuss the main definitions and semianalytical methods to evaluate single-bunch instabilities. Two possible instability mechanisms in the SPS and a comparison of calculations with measurements are presented in Sec.~\ref{sec: sps_instability}. We consider different instability mitigation measures in Sec.~\ref{sec: mitigation} and, finally, present the main conclusions.

\section{Main equations and definitions}\label{sec: main}

The longitudinal motion of a particle in a synchrotron can be described in terms of its energy and phase deviations, $\Delta E$ and $\phi$, relative to the synchronous particle with the energy $E_0$. For beam stability analysis it is convenient to introduce another set of variables, the energy $\mathcal{E}$ and the phase $\psi$ of synchrotron oscillations
\begin{align}
    \mathcal{E} &= \frac{\Dot{\phi}^2}{2\omega_{s0}^2} + U_t(\phi), \\
    \psi &= \operatorname{sgn}(\eta \Delta E)\frac{\omega_s(\mathcal{E})}{\sqrt{2}\omega_{s0}} \int_{\phi_{\text{max}}(\mathcal{E})}^{\phi} \frac{d\phi'}{\sqrt{\mathcal{E}-U_t(\phi')}}. \label{eq:psi}
\end{align}
Here  $\eta = 1/\gamma^2_\mathrm{tr} - 1/\gamma^2$ is the phase slip factor, $\gamma$ is the relativistic Lorentz factor, $\gamma_\mathrm{tr}$ is the Lorentz factor at transition energy, $f_{s0} = \omega_{s0}/2\pi$ is the frequency of small-amplitude synchrotron oscillations in a bare single-rf system,  $\omega_s(\mathcal{E})$ is the
synchrotron frequency as a function of $\E$, and  $\phi_{\text{max}} (\mathcal{E})$ is the maximum phase of the particle with synchrotron oscillation energy  $\mathcal{E}=U_t[\phi_{\text{max}} (\mathcal{E})]$. The total potential includes contributions from both the rf system and the beam-induced fields
\begin{equation}
    \label{eq:potential}
    U_t(\phi) = \frac{1}{V_1\cos{\phi_{s0}}}\int^\phi_{\Delta \phi_{s}}  \left[V_\mathrm{rf}(\phi^\prime) + V_\mathrm{ind}(\phi^\prime) - \delta E_0/q \right]\,  d\phi^\prime.
\end{equation}
where $V_1$ is the rf voltage amplitude of the main rf system, $\delta E_0$ is the energy gain per turn of the synchronous particle with charge $q$ excluding intensity effects, $\Delta \phi_s$ is the synchronous phase shift due to intensity effects that satisfies the relation $\delta E_0/q = V_1 \sin{\phi_{s0}} = V_\mathrm{rf}(\Delta \phi_s) + V_\mathrm{ind}(\Delta \phi_s) $, and $\phi_{s0}$ is the synchronous phase in a bare single-rf system. Below we consider a double-rf system with a total voltage of
\begin{equation}
    \label{eq:drf}
    V_\mathrm{rf}(\phi) = V_1\left[\sin{\left(\phi + \phi_{s0} \right)} + r \sin{\left(n\phi + n\phi_{s0} + \Phi_n \right)}\right],
\end{equation}
where $\Phi_n$ is the relative phase offset between the main and the HH rf systems with harmonic numbers $h$ and $nh$, respectively, and  $V_{n} = r V_1$ is the voltage amplitude of the HH rf system.
For particular values of $\Phi_n$ one can define two distinct regimes: bunch-shortening mode~(BSM) when both rf systems are in phase at the bunch center and bunch-length mode (BLM) when both rf systems are in counter-phase at the bunch center. In the SPS operation, they are chosen such that the contribution of the HH rf system is zero at $\phi=0$ and it does not contribute to a shift of the synchronous phase. Thus, $\Phi_n=\pi-n\phi_{s0}$ for BSM and $\Phi_n=-n\phi_{s0}$ for BLM.

In general, the total potential $U_t$ depends on the particle distribution function, impedance model, and bunch intensity.
It can be calculated using an iterative procedure~\cite{Dyachkov:1995, burov2012van}.
In this work, we consider a particle distribution of the binomial family 
\begin{equation}
    \label{eq:binom_distr}
    \F(\E) = \frac{1}{2\pi\omega_{s0}A_N}\left(1-\frac{\E}{\E_{\text{max}}}\right)^{\mu},
\end{equation}
with the normalization constant
\begin{equation}
    A_N = \omega_{s0}\int_0^{\E_{\text{max}}}\frac{d\E}{\omega_s(\E)}\left(1-\frac{\E}{\E_{\text{max}}}\right)^{\mu}.
\end{equation}
For $\mu \to \infty$, the bunch has a Gaussian line density and the corresponding bunch length $\tau_{4\sigma}$ is typically defined as four times the rms bunch length $\sigma$, i.e. $\tau_{4\sigma} = 4 \sigma$. The bunch length $\tau_{4\sigma}$ is related to the full-width at half-maximum (FWHM) bunch length $\tau_\mathrm{FWHM}$ as
\begin{equation}
   \label{eq:bunch_length}
   \tau_{4\sigma} = \tau_\mathrm{FWHM} \sqrt{2/\ln 2}.
\end{equation}
In practice, SPS proton bunches are far from being Gaussian, and the best fit to the measured bunch profiles is for $\mu \approx 1.5$. This value will be assumed in the present work for all calculations and simulations. For easy comparison with measurements, we use Eq.~(\ref{eq:bunch_length}) as a definition of the bunch length.
We also define the total longitudinal emittance in units of~eVs as
\begin{equation}
    \label{eq:emittance}
    \epsilon = \oint  \frac{\Delta E(\phi)}{h\omega_0} d\phi = \sqrt{-\frac{V_1 \cos\phi_{s0} \; q \beta^2 E_0}{\pi \eta \omega_0^2 h^3}} \epsilon_N,
\end{equation}
with a dimensionless emittance 
\begin{equation}
    \label{eq:norm_emittance}
    \epsilon_N = 2\int_{\phi_{\min}}^{\phi_{\max}}  \sqrt{(\mathcal{E}_{\max} - U_t(\phi))}\; d\phi,
\end{equation}
where $\omega_0= 2 \pi f_0$ is the revolution frequency, $\phi_\mathrm{min}$ and $\phi_\mathrm{max}$ are the minimum and maximum phases of the particle with the energy of synchrotron oscillation $\E_{\max}$.

\subsection{Linearized Vlasov equation}\label{subsec: Vlasov Equation}
When the perturbation $\Tilde{\mathcal{F}}$ to the stationary particle distribution function $\mathcal{F}(\E)$ 
grows with time $t$, the beam is unstable. The initial time evolution of $\Tilde{\mathcal{F}}$ is dictated by the linearized Vlasov equation (e.g.~\cite{AChao1993})
\begin{equation}
    \label{eq:lin_vl_eq}
    \frac{\partial \Tilde{\mathcal{F}}}{\partial t} + \frac{d \mathcal{F}}{d\mathcal{E}}\frac{d\mathcal{E}}{dt} + \frac{\partial \Tilde{\mathcal{F}}}{\partial \psi} \frac{d\psi}{dt} = 0.
\end{equation}
After expansion over the azimuthal harmonics $m$ of synchrotron motion, the solution of Eq.~(\ref{eq:lin_vl_eq}) at frequency $\Omega$ with the eigenfunctions $C_m(\mathcal{E},\Omega)$~\cite{OY:1990}
\begin{align}
    &\Tilde{\mathcal{F}}(\mathcal{E},\psi,t) \nonumber \\&= e^{-i\Omega t}\sum_{m=1}^\infty C_m(\mathcal{E},\Omega)\left[\cos m\psi + \frac{i\Omega}{m\omega_s(\E)}\sin m\psi\right]
\end{align}
leads to the integral equation
\begin{align}
    \label{eq:integral_eq_OY}
    \left[\Omega^2 \right. &- \left. m^2\omega^2_s(\E) \right]C_m(\E,\Omega)  \nonumber\\ & =2 i \zeta \omega^2_{s0}\;   m^2 \omega^2_s(\E) \frac{d\F(\E)}{d\E} 
    \sum_{m^\prime =1 }^{\infty} \int_0^{\E_{\max}}\frac{d\E^\prime}{\omega_s(\E^\prime)}\nonumber\\ &\times \sum_{k=-\infty}^{\infty}\frac{Z_k(\Omega)/k}{hZ_0} I_{mk}(\E) I^*_{m^\prime k}(\E^\prime) C_{m^\prime}(\E^\prime,\Omega).
\end{align}
Here $Z_k(\Omega)=Z(k \omega_0 + \Omega)$ is the longitudinal impedance at frequency $k \omega_0 + \Omega$ and  $Z_0\approx 377\; \Omega$ is the impedance of free space. We also introduced the dimensionless intensity parameter 
\begin{equation}\label{eq:xi}
    \zeta = \frac{q N_p h^2 \omega_0 Z_0}{V_1\cos\phi_{s0}}
\end{equation}
with $N_p$ being the bunch intensity. The function $I_{mk}(\mathcal{E})$ is defined as
\begin{equation}
    \label{eq: Imk}
    I_{mk}(\mathcal{E}) = \frac{1}{\pi}\int_0^\pi  e^{ik\phi(\mathcal{E},\psi)/h}\cos m\psi \; d\psi.
\end{equation}
Detailed derivations of Eq.~(\ref{eq:integral_eq_OY}) using variables ($\mathcal{E},\psi$) can be found in~\cite{Karpov:2021} and it is identical to Eq.~(41) therein.
For a given impedance model, the single-bunch stability depends on the two dimensionless parameters, $\zeta$ and $\epsilon_N$, as well as on the parameters of the rf potential: $n$, $\phi_{s0}$, $r$, and $\Phi_n$. This fact will be applied below to understand the  mechanisms of instabilities observed in the SPS.

\subsection{van Kampen modes}\label{subsec: vK modes}

The integral equation (\ref{eq:integral_eq_OY}) is equivalent to the equation that describes collective modes in a plasma~\cite{vKampen1, vKampen2, Case1959}. An initial perturbation can be expressed as a superposition of van Kampen modes, which are, in general, described by nonregular functions. To show this, one can first perform the substitution in Eq.~(\ref{eq:integral_eq_OY})
\begin{equation}
    \label{eq: substitution}
    C_m(\E,\Omega) = \sqrt{-\omega_s(\E)\frac{d\F(\E)}{d\E}}m\frac{\omega_s(\E)}{\omega^2_{s0}}\Tilde{C}_m(\E,\Omega),
\end{equation}
which leads to
\begin{align}
    \label{eq: symmetric OY}
    \left[\frac{\Omega^2}{\omega^2_{s0}} - \frac{m^2 \omega^2_s(\E)}{\omega^2_{s0}} \right]&\Tilde{C}_m(\E, \Omega)  \nonumber \\ =-2i\zeta \sum_{m^\prime=1}^{\infty}& \int_{0}^{\E_{\max}} K_{m m^\prime}(\E,\E^\prime) \Tilde{C}_{m^\prime,\Omega}(\E^\prime)d\E^\prime.
\end{align}
Here, the kernel $K$ is defined as
\begin{align}
    \label{eq: symm kernel}
    K_{mm^\prime}(\E,\E^\prime,\Omega) &= \sum_{k=-\infty}^{\infty} \frac{Z_k(\Omega)/k}{h Z_0}  \nonumber\\ & \times m\sqrt{-\omega_s(\E)\frac{d\F(\E)}{d\E}} I_{mk}(\E) \nonumber \\
    &\times m^\prime \sqrt{-\omega_s(\E^\prime)\frac{d\F(\E^\prime)}{d\E^\prime}} I^*_{m^\prime k}(\E^\prime).
\end{align}
Introducing a set of orthonormal functions $s_n^{(m)}$,
\[
\int_0^{\E_{\max}} s^{(m)}_n(\E)\; s^{(m)}_{n^\prime} (\E) d\E = \delta_{n n^\prime},
\]
with the Kronecker delta $\delta_{ij}$, we can decompose $\Tilde{C}_m$ and $K_{mm^\prime}$, similarly to~\cite{YHChin:1983}:
\begin{align}
    \Tilde{C}_m\left(\E,\Omega\right) &= \sum_{n=0}^\infty a_m^n (\Omega) s^{(m)}_n (\E), \label{eq: polinom expansion C}\\
    K_{mm^\prime}(\E,\E^\prime,\Omega) &= \sum_{n=0}^\infty \sum_{n^\prime = 0}^\infty K^{n n^\prime}_{m m^\prime} s^{(m)}_n (\E) s^{(m^\prime)}_{n^\prime} (\E^\prime).\label{eq: polinom expansion K}
\end{align}
with the coefficients $a_m^n$ and $K^{n n^\prime}_{m m^\prime}$ defined as
\begin{equation} 
    \label{eq: expansion coeff a}
    a_m^n(\Omega) = \int_0^{\E_{\max}} \Tilde{C}_m\left(\E,\Omega\right) s^{(m)}_n (\E) d\E,
\end{equation}
and
\begin{align}
    & K^{n n^\prime}_{m m^\prime}(\Omega)
\nonumber \\ &=\int_0^{\E_{\max}} \int_0^{\E_{\max}} K_{mm^\prime}(\E,\E^\prime,\Omega) s^{(m)}_n (\E) s^{(m^\prime)}_{n^\prime} (\E^\prime) d\E d\E^\prime, \label{eq: expansion coeff K}
\end{align}
respectively.
Finally, inserting Eqs.~(\ref{eq: polinom expansion C}, \ref{eq: polinom expansion K}) in Eq.~(\ref{eq: symmetric OY}), we obtain the eigenfunctions 
\begin{align}
    \label{eq: van Kampen modes}
    &\Tilde{C}_m\left(\E,\Omega\right) \nonumber \\ &=-\left\{\text{P}\frac{1}{\Omega^2 - m^2 \omega^2_s(\E)} + \alpha(\E) \delta\left[\Omega^2 - m^2 \omega^2_s(\E) \right]\right\} \nonumber\\
    &\times 2i\zeta \omega^2_{s0}\sum_{m^\prime=1}^\infty \sum_{n=0}^\infty \sum_{n^\prime=0}^\infty K_{mm^\prime}^{nn^\prime}(\Omega)a_{m^\prime}^{n^\prime} (\Omega) s^{(m)}_n(\E),
\end{align}
where P denotes the principal value of the integral, and $\alpha(\E,\Omega)$ can be found from the normalization condition 
\[\sum_{m=1}^\infty \int_0^{\E_{\max}}\Tilde{C}_m(\E, \Omega) d\E = 1.
\]
For Im$\Omega\to \pm0$, for example, one gets $\alpha=\mp i\pi$. 

Most of the stable modes are located within the incoherent spectrum $\Omega \in m\omega_s(\E)$ and have singular eigenfunctions due to the second term in braces in Eq.~(\ref{eq: van Kampen modes}). 
Unstable modes and modes for which Landau damping is lost, $\Omega \not \in m\omega_s(\E)$, have regular eigenfunctions as they are defined by the first terms in braces.  

\subsection{Instability mechanisms\label{subsec: instability mechanisms}}

The eigenfunctions are related to the perturbed line density harmonics (see, e.g. in~\cite{Karpov:2021})
\begin{equation}
    \label{eq:line_density_pert_harm_OY}
    \Tilde{\lambda}_k (\Omega) = \frac{\omega^2_{s0}}{h} \sum_{m =1 }^{\infty} \int_0^{\E_{\max}}  \frac{C_m(\E,\Omega) I^*_{mk}(\E)}{\omega_s(\E)}d\E. 
\end{equation}
Multiplying Eq.~(\ref{eq: symmetric OY}) with $\Tilde{C}^*_m(\E,\Omega)$ and integrating over $\E$ leads to
\begin{align}
    \label{eq: mode frequency}
    \Omega^2 &= \sum_{m=1}^\infty \int_0^{\E_{\max}}m^2 \omega^2_s(\E) \frac{|\Tilde{C}_m(\E, \Omega)|^2}{B(\Omega)} d\E 
    \nonumber \\
    &+ \frac{2\zeta h \omega^2_{s0}}{B(\Omega)} \sum_{k=-\infty}^{\infty} \frac{ \text{Im}Z_k(\Omega)/k}{Z_0} \left | \Tilde{\lambda}_k(\Omega) \right |^2
    \nonumber \\
    &- \frac{2i\zeta h \omega^2_{s0}}{B(\Omega)} \sum_{k=-\infty}^{\infty} \frac{ \text{Re}Z_k(\Omega)/k}{Z_0} \left | \Tilde{\lambda}_k(\Omega) \right |^2,
\end{align}
where
\[ B(\Omega)=\sum_{m=1}^\infty\int_0^{\E_{\max}}|\Tilde{C}_m(\E, \Omega)|^2 d\E.
\]
Below the instability threshold, i.e. Im$\Omega=0$, the contribution of the last term in Eq.~(\ref{eq: mode frequency}) must be zero as it is purely imaginary.
Depending on the sign of the sum on the second line of Eq.~(\ref{eq: mode frequency}), the mode frequency can be shifted upwards or downwards with respect to the weighted incoherent frequency, i.e. first line of Eq.~(\ref{eq: mode frequency}). Thus, two modes which frequencies approach each other can couple leading to the mode-coupling instability. In some cases one can distinguish two types of instabilities: radial and azimuthal mode-coupling. The former usually appears when different azimuthal modes are well separated as a result of a small synchrotron frequency spread and two coupled modes belong to the same azimuthal mode. The latter mechanism requires significant frequency shift of the modes as well as significant synchrotron frequency spread, so the modes with different azimuthal mode numbers can couple. Both instability types can be found using self-consistent analysis. Neglecting PWD and rf nonlinearity only the classical azimuthal mode-coupling instability~\cite{Sacherer:1977} is possible (see Appendix~\ref{annex}).

In the case of instability, we obtain from Eq.~(\ref{eq: mode frequency}) the growth rate
\begin{equation}
    \label{eq:growth_rate}
    \text{Im}\Omega =- \frac{\zeta h \omega^2_{s0}}{B(\Omega)\text{Re}\Omega} \sum_{k=-\infty}^{\infty} \frac{\text{Re}Z_k(\Omega)/k}{Z_0} \left|\Tilde{\lambda}_{k}(\Omega)\right|^2,
\end{equation}
which shows that for a smooth impedance, $Z_{-k}(\Omega)\approx Z^*_{k}(\Omega)$, the spectrum of unstable modes must be an asymmetric function of frequency $kf_0$. A similar conclusion was obtained in~\cite{Mosnier:1999}, pointing out the possibility of a radial mode-coupling instability within a single azimuthal mode in the presence of PWD. 
Equation~(\ref{eq:growth_rate}) shows, however, that the asymmetry of the mode spectrum is the general property of unstable modes, irrespective of the single-bunch instability mechanism.

Below we will discuss the methods of computing the van Kampen modes for cases of practical applications.
\subsection{Methods to solve the linearized Vlasov equation \label{subsec: Numerical methods}}

To solve the integral equation~(\ref{eq:integral_eq_OY}), it is usually converted to the infinite system of equations
\begin{equation}
    \label{eq: orthogonal system}
    \frac{\Omega^2}{\omega^2_{s0}}a^n_m (\Omega)= \sum_{n^\prime = 0}^\infty \sum_{m^\prime=1}^\infty M^{n n^\prime}_{m m^\prime}(\Omega) a^{n^\prime}_{m^\prime} (\Omega),
\end{equation}
where the matrix elements are defined as (see, e.g.~\cite{Besnier:1979,YHChin:1983})
\begin{align}
    \label{eq: matrix orthogonal}
    M^{n n^\prime}_{m m^\prime}(\Omega) &= m^2\delta_{m m^\prime} \int_0^{\E_{\max}}\frac{\omega^2_s(\E)}{\omega^2_{s0}}s^{(m)}_n(\E) s^{(m)}_{n^\prime}(\E)d\E \nonumber \\
    &- 2i\zeta K_{m m^\prime}^{n n^\prime}(\Omega).
\end{align}
The system of equations~(\ref{eq: orthogonal system}) becomes linear, if the dependence of the matrix elements $M^{nn^{\prime}}_{mm^\prime}$ on $\Omega$ can be neglected. This requires a rather smooth dependence of impedance on the frequency, so that $Z_k(\Omega)\approx Z_k(0)$, which is valid for the SPS impedance where the bandwidth of all relevant resonant peaks is larger than the revolution frequency $f_0$.
Then, after truncation of the infinite sums over indices $m$, $n$, and $k$,  Eq.~(\ref{eq: orthogonal system}) can be solved as a standard eigenvalue problem.

Two methods exist to compute the matrix elements depending on the choice of the orthonormal functions: the orthogonal polynomial expansion, and the Oide-Yokoya method~\cite{OY:1990}.
For the former method, the orthonormal functions for a particle distribution according to Eq.~(\ref{eq:binom_distr}) can be constructed from the Jacobi polynomials $P^{(\alpha,\beta)}_n(x)$, 
\begin{align}
\label{eq: orthogonal functions}
s^{(m)}_n(\E) &= \sqrt{\frac{ (2n+m+\mu) \Gamma(n+1)\Gamma(n+m+\mu) }{ \Gamma(n+m+1) \Gamma(n+\mu)}} \nonumber \\
&\times \sqrt{-\frac{2\pi \omega_{s0} A_N}{\mu}\frac{d\F(\E)}{d\E}} \left(\frac{\E}{\E_{\max}}\right)^{m/2} \nonumber \\
&\times P^{(m,\mu-1)}_n\left(1-\frac{2\E}{\E_{\max}}\right),
\end{align}
where $n$ is the radial mode number, and $\Gamma(x)$ is the Gamma function.
For the latter approach step-like functions~\cite{OY:1990} are used 
\[
s^{(m)}_n (\E)= \left\{
  \begin{array}{ll}
     \frac{1}{\sqrt{\Delta \E_n}}, &  \E_n - \frac{\Delta \E_n}{2} < \E \leq \E_n + \frac{\Delta \E_n}{2} \\
    0, & \text{elsewhere,}
  \end{array}\right.
\]
where $\E_n$ is the $n$th mesh point on the energy grid, and $\Delta \E_n$ is the thickness of the corresponding strip. 
Additionally, integration is approximated by a sum, so the matrix elements of the eigenvalue problem~(\ref{eq: orthogonal system}) become
\begin{align}
    \label{eq:matrix}
    M^{n n^\prime}_{mm^\prime}(\Omega) &= \frac{m^2\omega_s^2\left(\E_n\right)}{\omega^2_{s0}}\delta_{nn^\prime}\delta_{mm^\prime} \nonumber \\ 
    & -2i\zeta \sqrt{\Delta \E_n \Delta \E_{n^\prime}} K_{mm^\prime}\left(\E_n,\E_{n^\prime},\Omega\right).
\end{align}

Both methods to solve the linearized Vlasov equation are implemented in the {\footnotesize MELODY} code and allow the evaluation of single-bunch stability for arbitrary impedance models. 
We find that the Oide-Yokoya method has better convergence properties than the orthogonal polynomial expansion. This is thanks to a nonuniform mesh with respect to the energy of synchrotron oscillations $\E_n$, which allows to improve the resolution around `critical' points. For example, in the case of instability with Re$\Omega = m\omega_s(\Tilde{\E})$, one can expect a resonance with a characteristic width of
\[
\delta \E \approx \left|\text{Im}\Omega \bigg/ \left.m\frac{d\omega_s(\E)}{d\E}\right|_{\E=\Tilde{\E}}\right|.
\] 
If $\delta \E \ll \E_{\max}$, a very high-resolution mesh in $\E$ might be needed to find a converged solution.

\section{Longitudinal single-bunch instabilities in the SPS}\label{sec: sps_instability}
In this section, we will show first the results of stability analysis at the SPS flat top (450~GeV) and then through acceleration.
The SPS has a double-rf system with frequencies of 200~MHz and 800~MHz ($n=4$). The main accelerator and beam parameters are listed in Table~\ref{tab: SPS_parameters}. Here, we will discuss a single-rf operation, while stability in a double-rf case will be addressed in Sec.~\ref{sec: mitigation}. 

It has been shown in previous studies that in a single-rf operation, the single-bunch intensity threshold is a noncontinuous and nonmonotonic function of the bunch length~\cite{Radvilas:2015, Lasheen:2017}. Since then, the SPS impedance model has been further refined, and below we will show the simulation results based on this latest model (Fig.~\ref{fig:sps_impedance}).
\begingroup
\begin{ruledtabular}
\begin{table}[b!]
	\caption{The SPS parameters at the flat-top energy~\cite{LHCDR3}. 
	}
	\begin{center}
		\begin{tabular}{l  c  c  c }
			Parameter & Units &  Value \\
			\hline
			Circumference, $C$ & m & 6911.554\\
			Beam energy, $E_0$ &GeV & 450 \\
			Transition Lorentz factor, $\gamma_\mathrm{tr}$ &  & 17.951 \\
			Main harmonic number, $h$&   & 4620 \\
			Main rf frequency, $f_\mathrm{rf}$& MHz & 200.394 \\
			Main rf voltage amplitude, $V_1$ & MV  & 7.2\\
		\end{tabular}
	\end{center}
	
	\label{tab: SPS_parameters}
\end{table}
\end{ruledtabular}
\endgroup

\begin{figure}[htp]
\begin{center}
\includegraphics{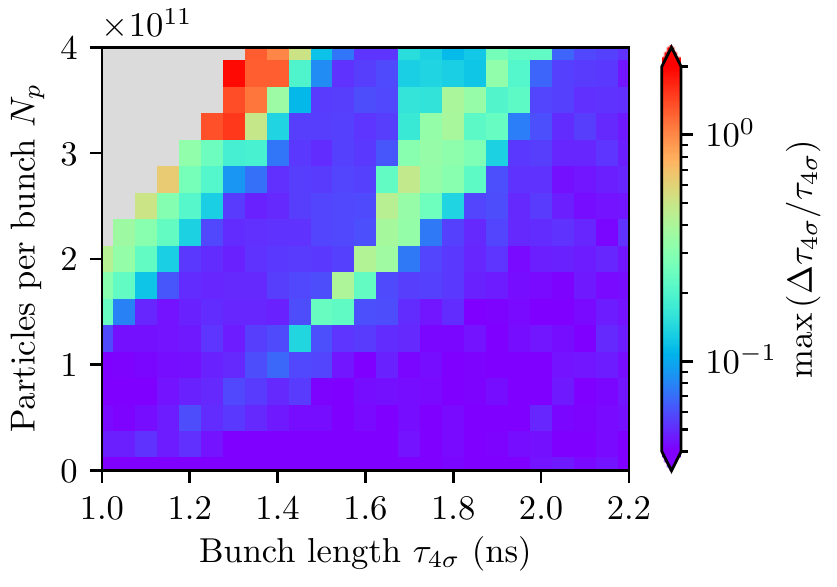}
\caption{Longitudinal stability map obtained from {\footnotesize BLonD} simulations for single-rf operation ($r=0$) and with the pre-LS2 impedance model (Fig.~\ref{fig:sps_impedance}). The color code indicates the maximum relative oscillation amplitude of the bunch length during $10^5$ turns. Gray area indicates parameters for which the initial stationary distribution was not found. The SPS parameters are according to Table~\ref{tab: SPS_parameters}. }
\label{fig:blond_only}
\end{center}
\end{figure}

The beam stability in simulations is probed by observing the growing oscillations of the bunch position (dipole) and length (quadrupole). Even though the actual evolution of the bunch parameters depends on the initial seed at the moment of bunch generation, a fast-growing instability can still be detected. 
In Fig.~\ref{fig:blond_only}, we show a stability map obtained from a scan in bunch intensity $N_p$~(in steps of $2.5\times 10^{10}$) and bunch length $\tau_{4\sigma}$~(in steps of 50 ps) in simulations for $10^6$ macroparticles. The initial bunches are matched to the total potential $U_t$ including intensity effects and then tracked for $10^5$ turns (about 600 synchrotron periods). The impedance model before the second long shutdown (LS2) shown in Fig.~\ref{fig:sps_impedance} was taken. 
In simulations, the maximum amplitude of the bunch-length oscillations $\Delta \tau_{4\sigma}$ divided by the average bunch length $\tau_{4\sigma}$ indicates the bunch stability~\cite{KarpovGadioux2021}.

\begin{figure}[htp]
\begin{center}
\includegraphics{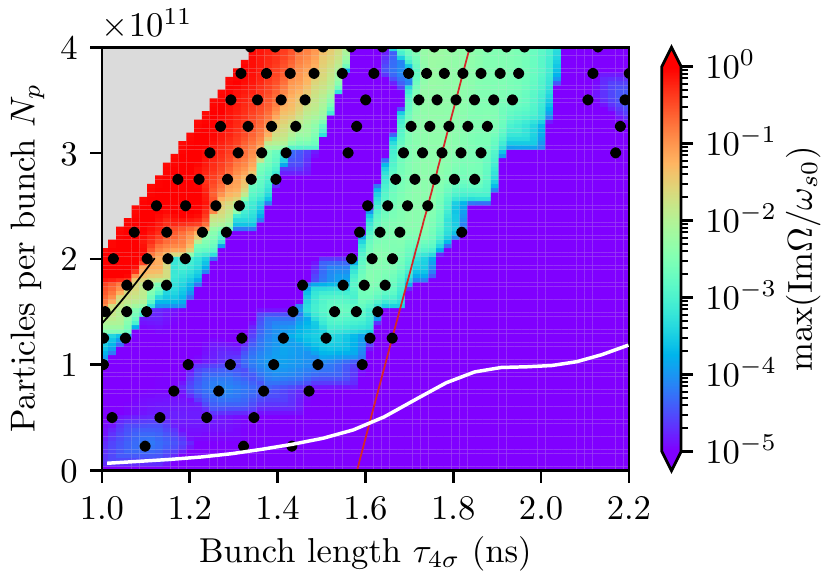}
\caption{Longitudinal stability map obtained from calculations with code {\footnotesize MELODY} (bottom) for the same parameters as in Fig~\ref{fig:blond_only}. The color code shows the growth rate of the most unstable mode, while the gray area indicates parameters for which stationary distribution was not found. The bunch parameters for which synchrotron frequency is a nonmonotonic function of $\E$ are shown as black circles. The black and red lines show examples of the bunch-length dependence on intensity due to PWD for constant energy of synchrotron oscillations $\E_{\max}=0.12$ and $\E_{\max}=0.57$, respectively.
The solid white curve corresponds to the LLD threshold computed with {\footnotesize MELODY}. }
\label{fig:melody_only}
\end{center}
\end{figure}

The semianalytical results calculated using code {\footnotesize MELODY} with up to ten azimuthal modes are shown in Fig.~\ref{fig:melody_only}. The scan was performed on a grid of $(\E,\zeta)$ values. The bunch length~$\tau_{4\sigma}$ and intensity~$N_p$ values were computed for each pair of parameters and then used to create an interpolated map of the corresponding growth rates. The impedance model is truncated at $k_{\max{}}f_0~\approx~6.4$ GHz, while the radial mesh is automatically refined depending on the bunch length (in some cases up to $N_\E\approx 10^3$ mesh points in the energy of synchrotron oscillations).  

In general, one can see that the stable solutions (purple) correspond to {\footnotesize BLonD} simulations where bunch length stays constant (the oscillation amplitude is below the noise level $\Delta \tau_{4\sigma}/\tau_{4\sigma}<0.04$), while $\tau_{4\sigma}$ grows for most of the unstable solutions with nonnegligible growth rates.
Both methods show that there is an unstable `island' approximately for intensities above $1.5\times 10^{11}$ and the bunch length in the range of 1.5 to 1.9~ns.

In addition, in Fig.~\ref{fig:melody_only} the circles indicate the cases with a nonmonotonic behavior of the synchrotron frequency as a function of the energy of synchrotron oscillations. In the past, it was shown that coupling of radial modes within a single azimuthal can occur in this case~\cite{Oide:1995}. We observe that the instability often is correlated with the presence of $d\omega_s/d\E = 0$ inside the bunch, however, it is not always the case. This can be understood from the fact that for a particular set of bunch parameters the growth rate can be very small, and it is difficult to detect in simulations or to properly resolve in semianalytical calculations. Still, the nonmonotonic behavior of the synchrotron frequency distribution could be a useful indication that we can expect instability.

We also show in Fig.~\ref{fig:melody_only} with a solid white line, the threshold of loss of Landau damping~(LLD) computed for a dipole mode ($m=1$) in the same way as in~\cite{Karpov:2021}. The LLD threshold is a monotonic function, thus it increases for a larger bunch length. However, the actual instability threshold is significantly higher. Even though Landau damping is already lost and the coherent modes are outside the incoherent band, their amplitudes do not grow in simulations without external noise. In operation, the excitation of these modes,  for example, due to some noise in the rf system, is suppressed by the phase loop. Thus, we conclude that LLD does not trigger the observed instability of  a single-bunch in the SPS. Two other instability mechanisms will be discussed in the following subsection.

\subsection{Radial mode-coupling instability\label{subsec: radial MC}}
\begin{figure}[bt]
\begin{center}
\includegraphics{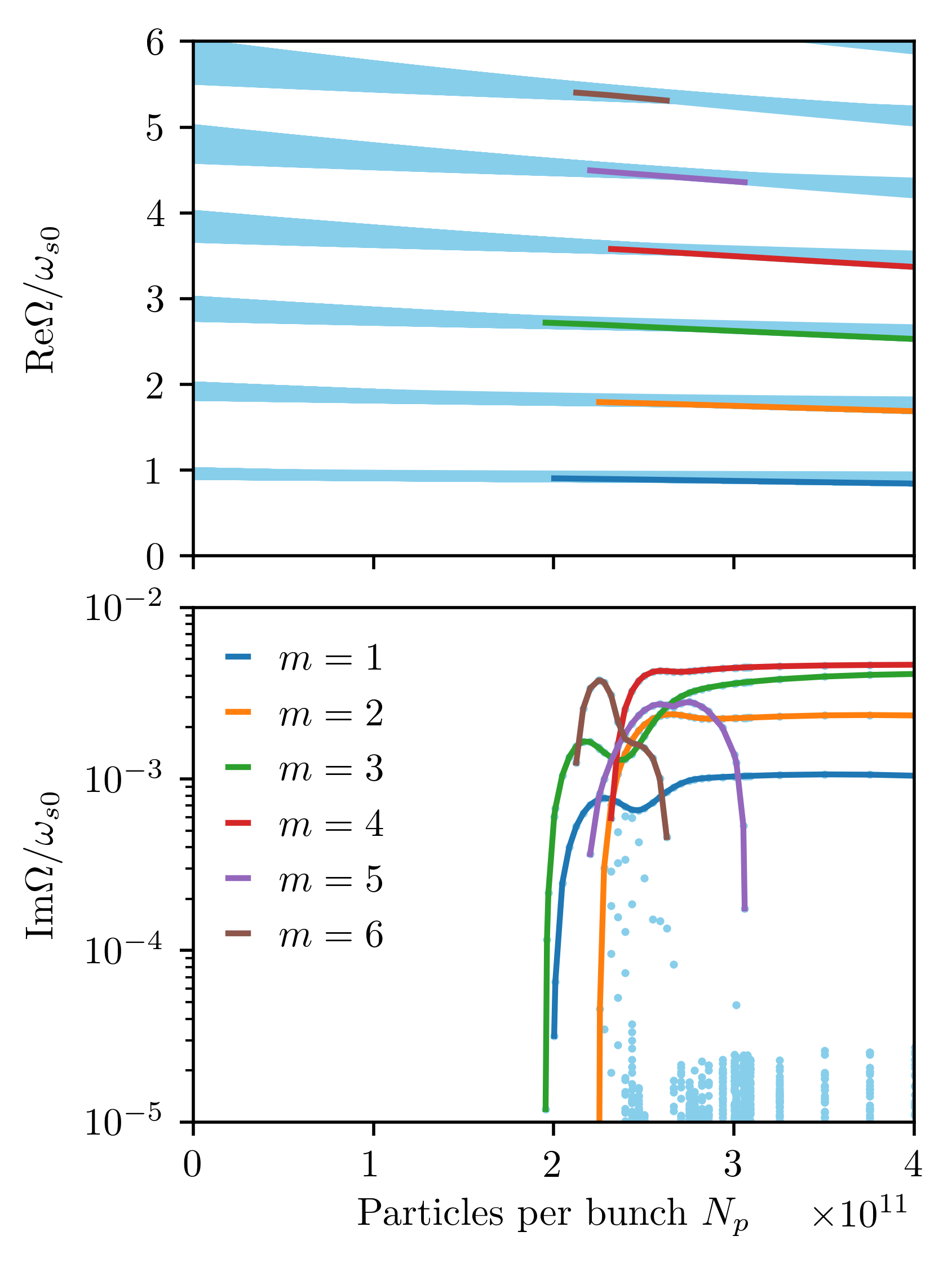}
\caption{Real (top) and imaginary (bottom) parts of van Kampen modes (light blue lines and dots) as a function of bunch intensity for $\E_{\max} = 0.57$. Unstable modes of the first six azimuthal modes are plotted as colored lines. Beam parameters correspond to those indicated by the red line in Fig.~\ref{fig:melody_only}.}
\label{fig: radial mode coupling}
\end{center}
\end{figure}
To clarify the instability mechanism inside an unstable `island' we plot van Kampen modes as a function of intensity for constant energy of synchrotron oscillations $\E_{\max}= 0.57$~(beam parameters according to the red line in Fig.~\ref{fig:melody_only}) in Fig.~\ref{fig: radial mode coupling}. The real part of the first six azimuthal modes (top plot) does not overlap for the entire intensity range. It means that only coupling of radial modes of the same azimuthal mode is possible for these parameters. The lowest threshold intensity is for the third azimuthal mode (green curve on the bottom plot of Fig.~\ref{fig: radial mode coupling}), while for higher intensity other azimuthal modes are also unstable. In that case, we can observe `microwave-like' instability as the bunch line density will be modulated by a mixture of several unstable modes. 

\begin{figure}[tb]
\begin{center}
\includegraphics{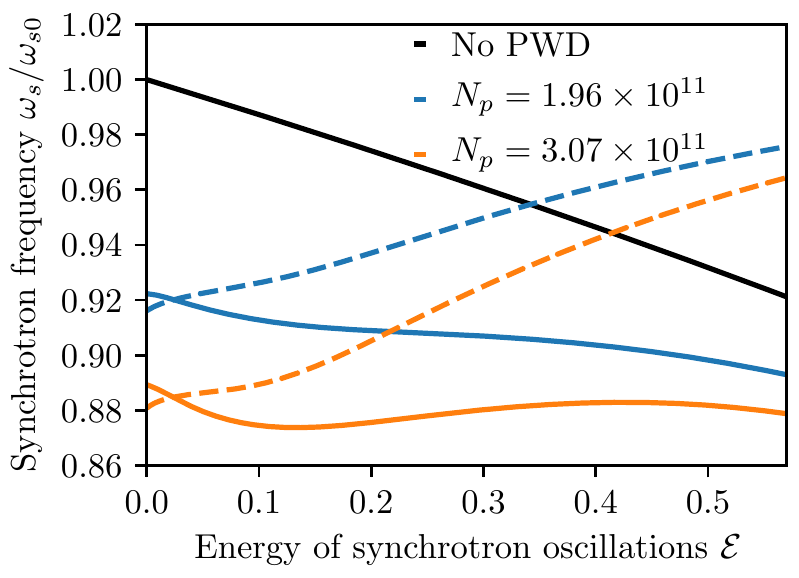}
\caption{Synchrotron frequency as a function of synchrotron-oscillation energy for different intensities and $\E_{\max}=0.57$. Solid lines are for single-rf potential, while dashed lines are for a parabolic-rf potential. The black curve illustrates the case without PWD.}
\label{fig: fs_distr_present2018}
\end{center}
\end{figure}

The synchrotron frequency distribution inside the unstable `island' is strongly affected by PWD and the initial rf nonlinearity plays an important role (see Fig.~\ref{fig: fs_distr_present2018}). At the threshold of instability (orange curve) $d\omega(\E)/d\E$ approaches zero and at the same time $d^2\omega(\E)/d\E^2=0$. This shape can be achieved for double-rf operation in BSM for $n>2$, while here it is a result of PWD. For even higher intensity the overlap of synchrotron frequencies for different $\E$ values becomes more significant.

\begin{figure}[bt]
\begin{center}
\includegraphics{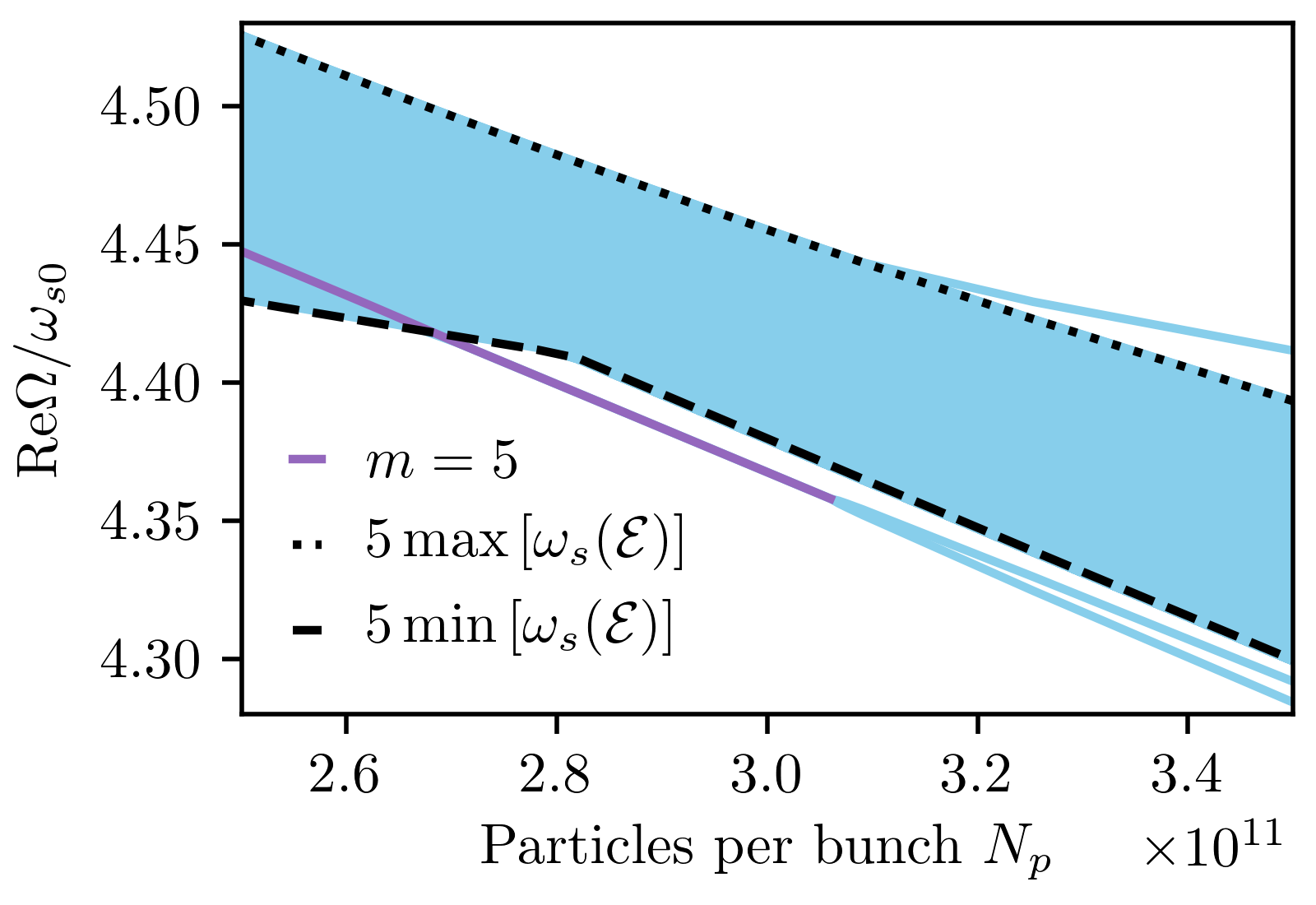}
\caption{Zoomed version of in Fig.~\ref{fig: radial mode coupling} (top plot). Blue lines represent van Kampen modes, while the unstable mode is shown in purple. The minimum and maximum incoherent synchrotron frequencies of the 5th azimuthal mode plotted as dashed and dotted lines, respectively.}
\label{fig: radial mode decoupling}
\end{center}
\end{figure}

In this example, instability occurs due to the coupling of two modes inside the incoherent synchrotron frequency spread. The unstable modes of the first four azimuthal modes move below the minimum incoherent frequency and remain unstable for higher intensities. The instability of higher-order azimuthal modes is suppressed above a certain intensity. For example, a zoomed region around the 5th azimuthal mode is shown in Fig.~\ref{fig: radial mode decoupling}. At the intensity of about $3.1\times10^{11}$ the modes are again decoupled, which results in suppression of this instability.

\begin{figure}[bt]
\begin{center}
\includegraphics{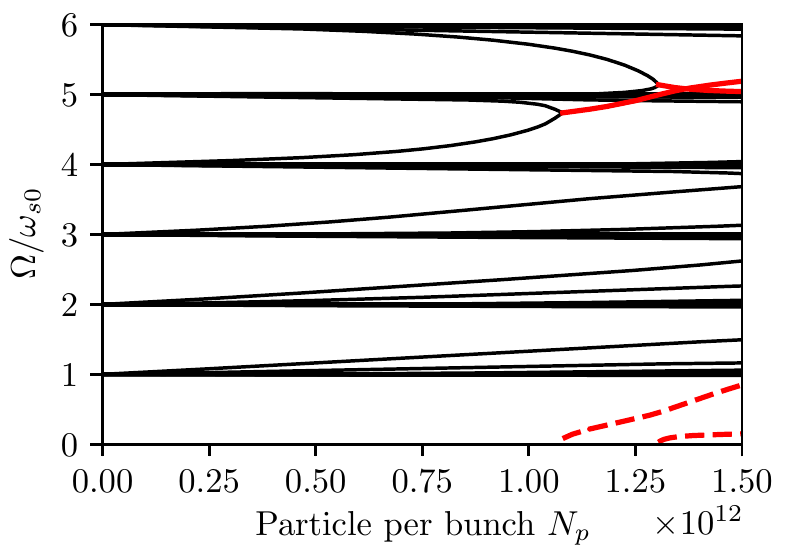}
\caption{Real (solid lines) and imaginary (dashed lines) parts of van Kampen modes as a function of bunch intensity for $\E_{\max} = 0.57$, neglecting rf nonlinearity and PWD.}
\label{fig: no spread 057}
\end{center}
\end{figure}

For completeness, the results of analysis when rf nonlinearity and PWD are neglected are also performed for the same $\E_{\max} =0.57$. As discussed in Sec.~\ref{subsec: instability mechanisms}, only azimuthal mode-coupling instability is possible in this case. Its threshold is almost a factor of 5 higher than the one obtained in self-consistent calculations (Fig.~\ref{fig: no spread 057}).  It means that the self-consistent analysis is crucial to predict the instability threshold for relatively long bunches.

\subsection{Azimuthal mode-coupling instability\label{subsec: azimuthal MC}}
\begin{figure}[bt]
\begin{center}
\includegraphics{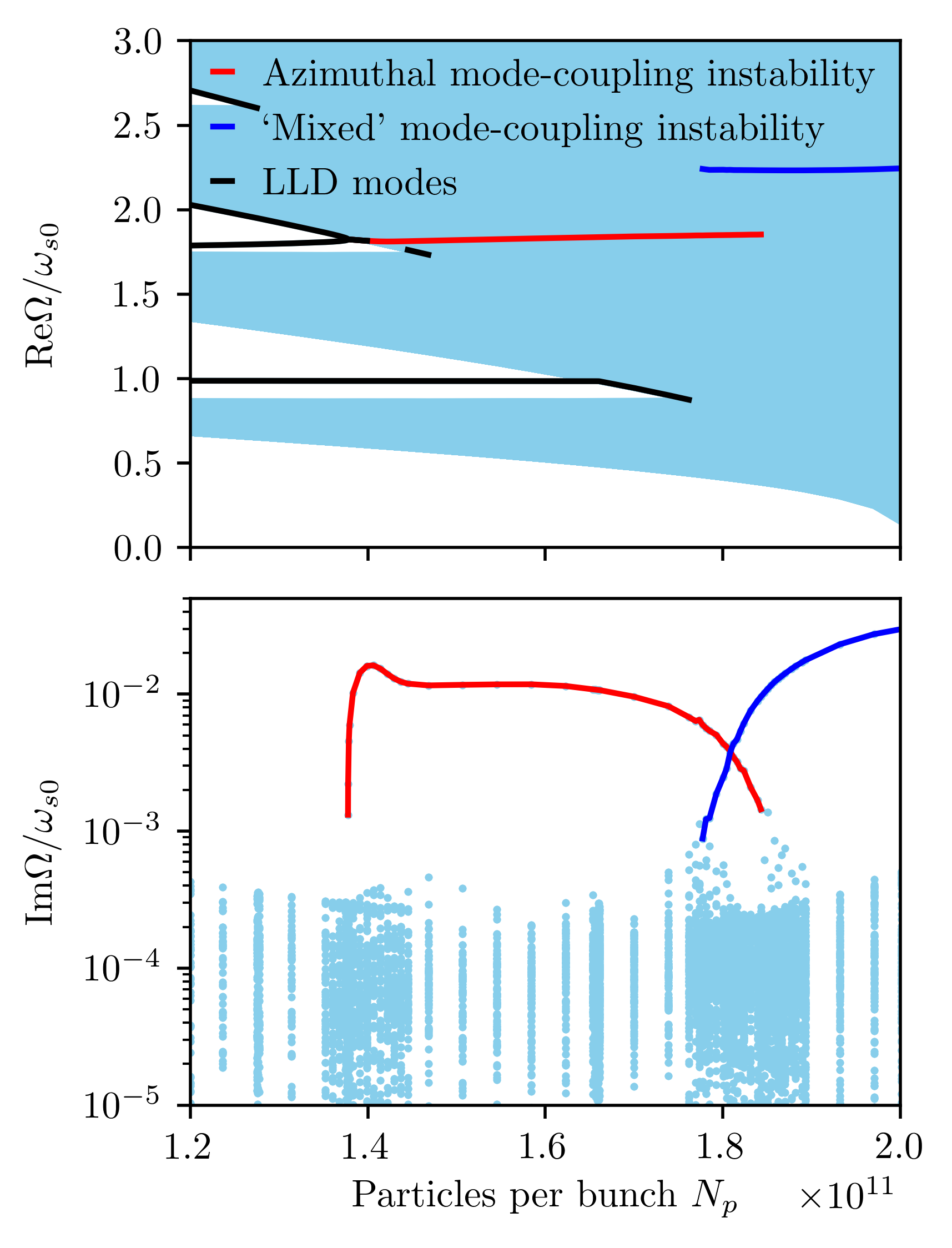}
\caption{Real (top) and imaginary (bottom) parts of van Kampen modes (blue lines and dots) as a function of bunch intensity for $\E_{\max} = 0.12$ (bunch parameters correspond to those indicated by the black line in Fig.~\ref{fig:melody_only}). Black lines show the frequencies of the mode for which Landau damping is lost. The red and blue lines indicate the  unstable modes.}
\label{fig: azimuthal mode coupling}
\end{center}
\end{figure}
\begin{figure}[tb]
\begin{center}
\includegraphics{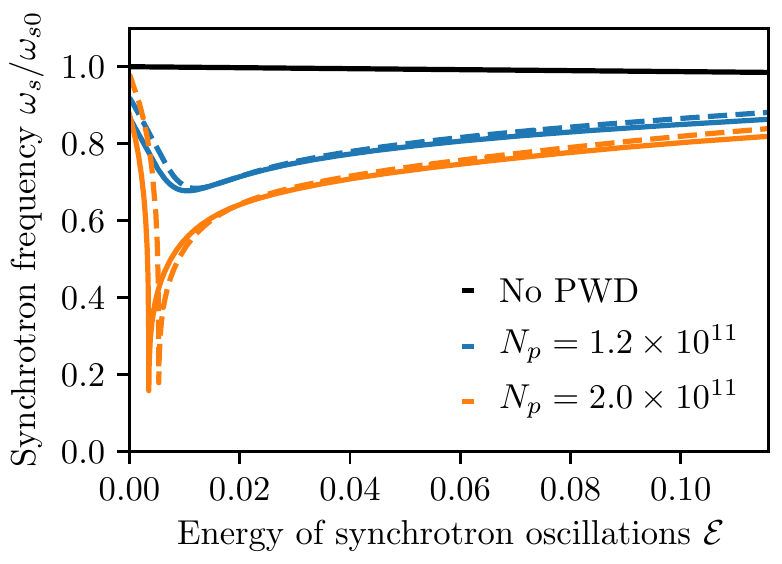}
\caption{Synchrotron frequency as a function of synchrotron-oscillation energy for different intensities and $\E_{\max}=0.12$. Solid lines are for single-rf potential, while dashed lines are for a parabolic-rf potential (linear rf voltage). The black curve illustrates to the case without PWD.}
\label{fig: fs_distr_present2018_azimuthal}
\end{center}
\end{figure}

A different mechanism of instability  can be observed for shorter bunches.
Figure~\ref{fig: azimuthal mode coupling} shows the real part of van Kampen modes of quadrupole and sextupole azimuthal modes as a function of intensity for $\E_{\max} = 0.12$ (black curve in Fig.~\ref{fig:melody_only}). One can see a quadrupole mode for which Landau damping is lost. It is moving towards the frequency band of radial modes with azimuthal mode number $m=3$.
\begin{figure}[bt]
\begin{center}
\includegraphics{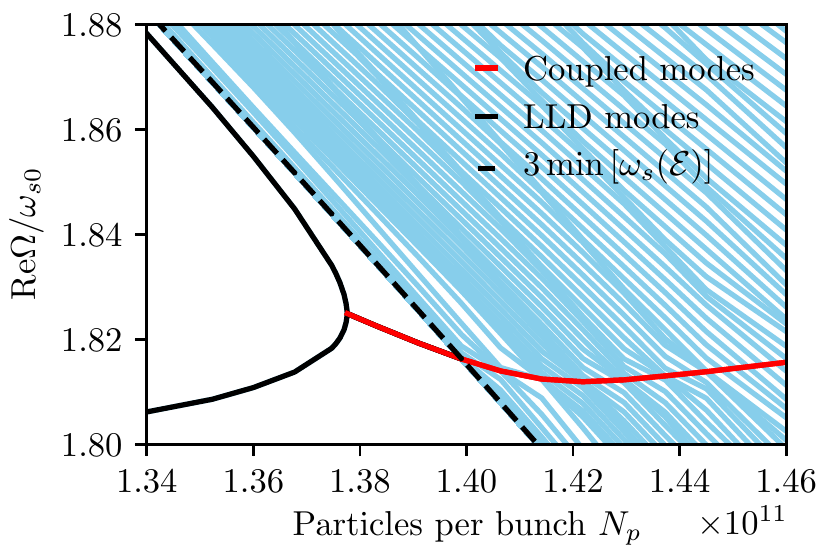}
\caption{Zoomed version of the top plot in Fig.~\ref{fig: azimuthal mode coupling} demonstrating the coupling of quadrupole and sextupole modes.
The dashed line is the minimum incoherent synchrotron frequency of the third azimuthal mode.}
\label{fig: azimuthal mode coupling zoom}
\end{center}
\end{figure}
At the same time, a sextupole mode of LLD type moves down towards the radial modes with $m=2$ (Fig.~\ref{fig: azimuthal mode coupling zoom}). 
This mode could emerge due to a strong PWD, which leads to the formation of a local minimum of the synchrotron frequency as a function of the synchrotron oscillation energy (blue curve in Fig.~\ref{fig: fs_distr_present2018_azimuthal}). Note also that rf nonlinearity does not play a significant role, as itself it results in a synchrotron frequency spread that is negligible in comparison to the one created by PWD (black curve in Fig.~\ref{fig: fs_distr_present2018_azimuthal}).
At the intensity of $1.37\times 10^{11}$, those two modes are coupled and thus result in azimuthal mode-coupling instability. The growth rate of the instability increases while the coupled modes remain outside the incoherent frequency bands ($N_p \lesssim 1.4\times 10^{11}$). Then, the growth rate is slightly suppressed and weakly depends on intensity once the coupled modes move inside the sextupole synchrotron frequency band. This is very different to the azimuthal-mode coupling instability predicted by Sacherer~\cite{Sacherer:1977} for which the growth rate is a strong function of intensity. Since this instability involves the LLD mode of the sextupole synchrotron frequency band, which emerges below the minimum synchrotron frequency, it is very sensitive to the particle distribution. As the distribution modifies after of the start of the instability, the modes can be decoupled again leading to beam stabilization.

Equation~(\ref{eq: mode frequency}) can be applied to understand why these modes move in different directions as a function of intensity. 
\begin{figure}[bt]
\begin{center}
\includegraphics{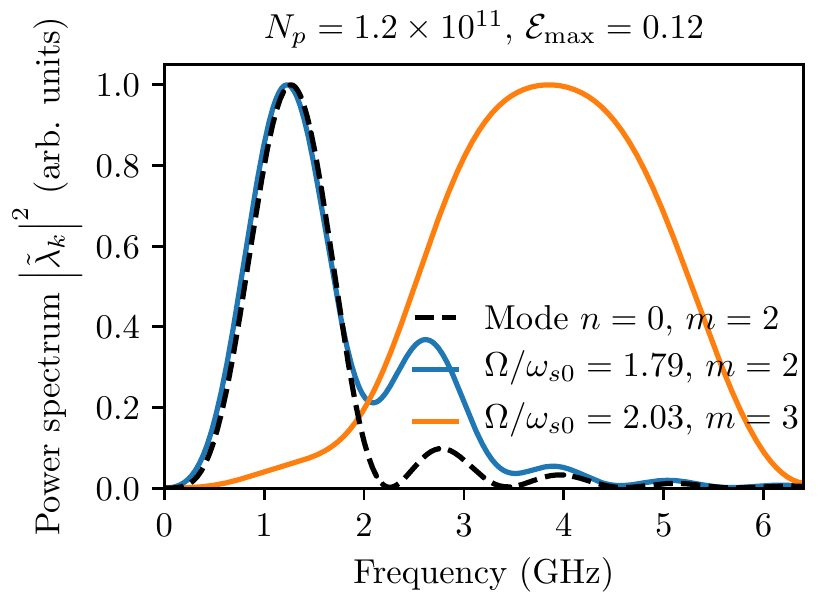}
\caption{Power spectra of quadrupole (blue) and sextupole (orange) azimuthal modes of the LLD type. The dashed line corresponds to the zeroth radial mode of the quadrupole azimuthal mode evaluated using Eqs.~(\ref{eq:line_density_pert_harm_OY})~and~(\ref{eq: orthogonal functions}).}
\label{fig: mode power spectra}
\end{center}
\end{figure}
Figure~\ref{fig: mode power spectra} shows that the power spectrum of the quadrupole mode (blue curve) has a larger overlap with positive values of $\text{Im}Z_k/k$ (Fig.~\ref{fig:sps_impedance}). Thus, its frequency shift will be positive as the cumulative sum $\sum_k \text{Im}Z_k/k \left|\Tilde{\lambda}_k \right|^2$ is positive. The opposite situation is true for the LLD mode with $m=3$, for which the maximum power spectrum is localized at the higher frequencies where $\text{Im}Z_k/k$ is mostly negative. In this case, the mode is moving downwards in frequency as the intensity increases.

\begin{figure}[bt]
\begin{center}
\includegraphics{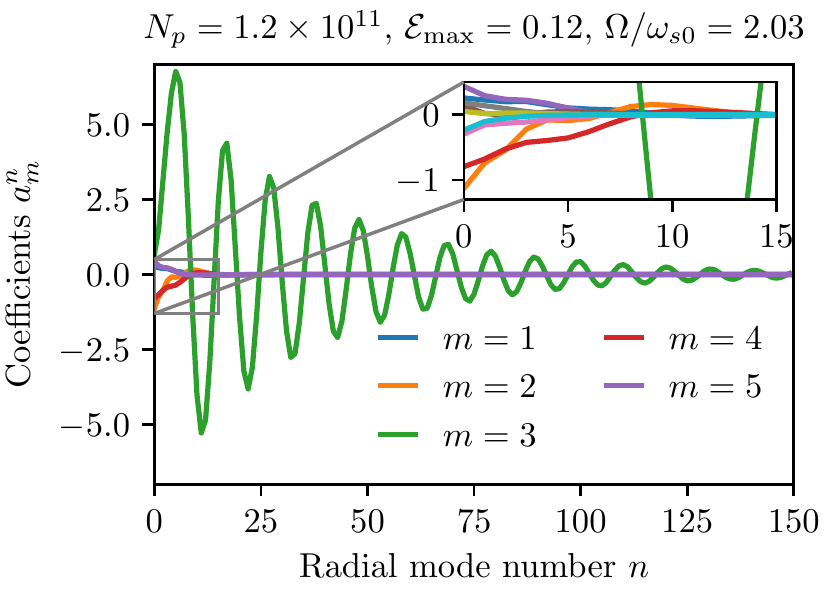}
\caption{Coefficients of the orthogonal polynomial expansion according to Eq.~(\ref{eq: expansion coeff a}) for the LLD mode shown as an orange curve in Fig.~\ref{fig: mode power spectra}.}
\label{fig: polynomial expansion example}
\end{center}
\end{figure}

We also find that the LLD mode above the maximum incoherent frequency of the second azimuthal mode is dominated by the zeroth radial mode (dashed curve in Fig.~\ref{fig: mode power spectra}), which is defined by Eqs.~(\ref{eq:line_density_pert_harm_OY})~and~(\ref{eq: orthogonal functions}). On the contrary, the LLD mode below the third azimuthal mode is actually a superposition of a few hundred radial modes, which makes it more difficult to find a converging solution using the orthogonal polynomial expansion. To see this precisely, the computed expansion coefficients $a^n_m$ from Eqs.~(\ref{eq: expansion coeff a})~and~(\ref{eq: orthogonal functions}) are shown in Fig.~\ref{fig: polynomial expansion example}.

There are no LLD modes above maximum incoherent frequency for $m>2$. This can be understood from the fact that their power spectra are localized at higher frequencies, and the modes sample mostly the negative values of $\text{Im}Z_k/k$. Thus, the modes are pushed inside the incoherent frequency band and remain Landau-damped.

Although the threshold of azimuthal mode-coupling instability is clearly defined ($N_p\approx1.37\times 10^{11}$), one can observe the modes with growth rates Im$\Omega/\omega_{s0} \sim 10^{-4}$ for lower intensities. In this case, $\omega_s(\E)$ has a local minimum~(Fig.~\ref{fig: fs_distr_present2018_azimuthal}) and, therefore, radial mode-coupling instability can appear. However, these unstable modes are localized at small values of $\E$, which makes it difficult to find converged solutions as was discussed in Sec.~\ref{subsec: Numerical methods}. Additionally, the spectra of these modes are localized at high frequencies and they sample a rather small residual part of the real impedance (Fig.~\ref{fig:sps_impedance}), which supports expectations of very small growth rates.

\begin{figure}[bt]
\begin{center}
\includegraphics{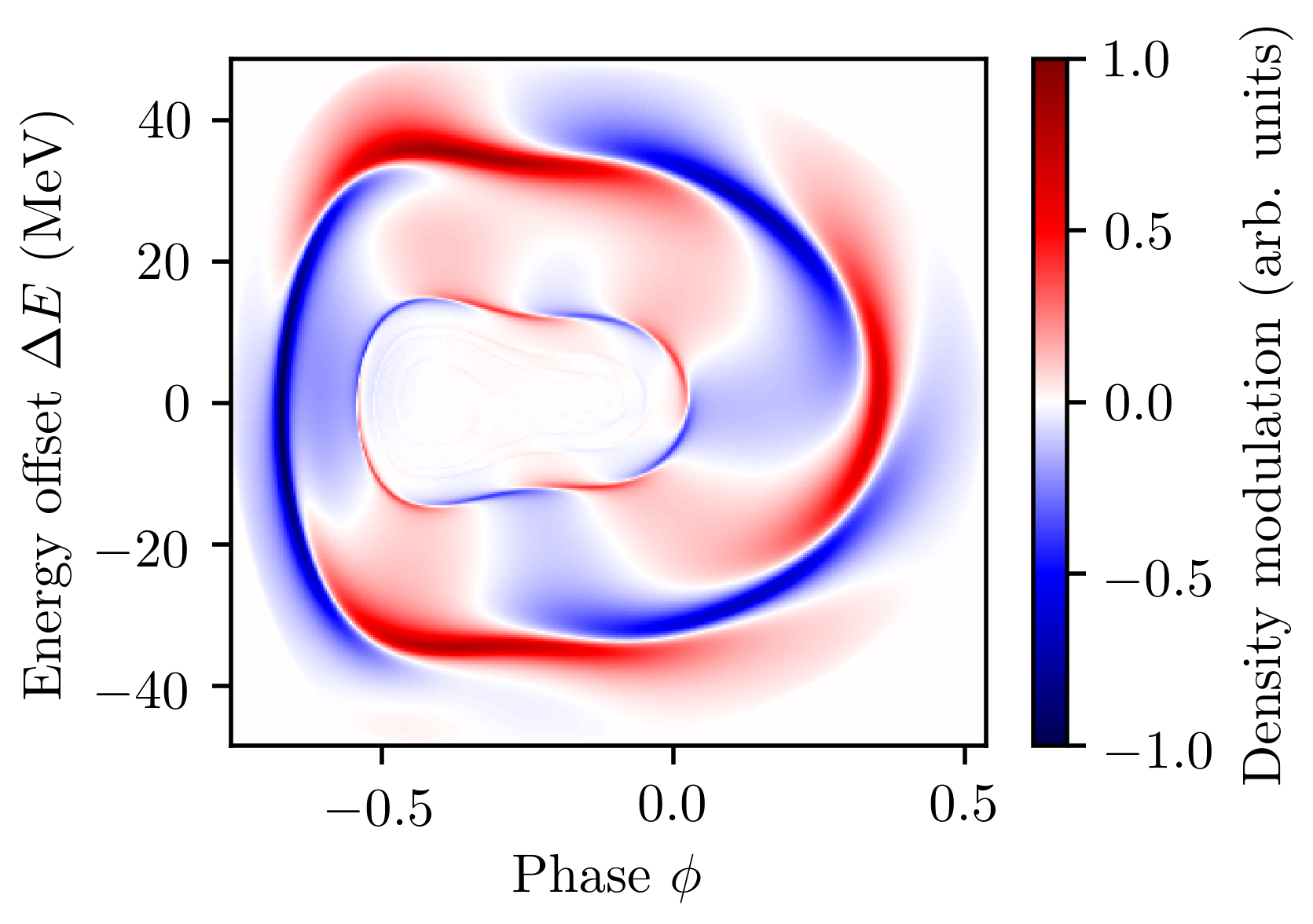}
\caption{Phase space of the most unstable mode for  $N_p=2.0\times 10^{11}$  and $\E_{\max{}} = 0.12$ computed with {\footnotesize MELODY} using the SPS impedance model from before LS2. The SPS parameters are according to Table~\ref{tab: SPS_parameters}.}
\label{fig:mode_phase_space}
\end{center}
\end{figure}

Above a certain intensity of about $\sim 1.8 \times 10^{11}$, we also observe another unstable mode (blue curve in Fig.~\ref{fig: azimuthal mode coupling}). It involves van Kampen modes, which are inside the synchrotron frequency spread leading to the radial mode-coupling instability. However, a strong PWD leads to an overlap of several azimuthal synchrotron frequency bands as the minimum synchrotron frequency is rather low (Fig.~\ref{fig: fs_distr_present2018_azimuthal}). The eigenvector of the unstable mode has resonant behavior for several azimuthal modes where $m\omega_s(\E) = \text{Re}\Omega$, which can seen in the longitudinal phase space (Fig.~\ref{fig:mode_phase_space}). This can be called a `mixed' mode-coupling instability, which has features of both radial and azimuthal mode-coupling instabilities.
For even higher intensities, the PWD leads to formation of the second local minimum of the potential well and the minimum synchrotron frequency approaches zero. In that case, there are two distinct branches for the possible solutions that start from two different fixed points~\cite{OY:1990, Cai2011}. In our calculations we show the results for the branch with the lowest minimum of the potential well and, in most cases, we observe an extremely fast instability with a growth time of a few synchrotron periods. Since the synchrotron frequency bands for different azimuthal modes significantly overlap, it is difficult to distinguish, which modes are coupled for a particular set of bunch parameters.

\begin{figure}[bt]
\begin{center}
\includegraphics{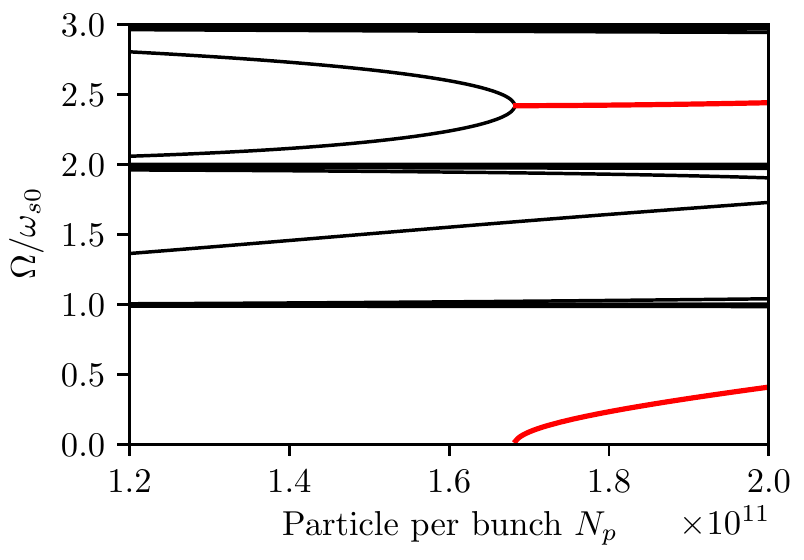}
\caption{Real (solid lines) and imaginary (dashed lines) parts of van Kampen modes as a function of bunch intensity for $\E_{\max} = 0.12$, when rf nonlinearity and PWD are neglected.}
\label{fig: no spread 013}
\end{center}
\end{figure}

Figure~\ref{fig: no spread 013} shows the results of calculations where rf nonlinearity and PWD are neglected, similarly to Fig.~\ref{fig: no spread 057} in the previous subsection. In that case, we again find the coupling of the quadrupole and sextupole azimuthal modes, while the instability threshold is about~25\% higher in comparison to the results of the self-consistent analysis. A similar dependence of the mode frequencies as a function of intensity can be achieved by using a single broad-band resonator impedance model with the resonant frequency at about 1.4 GHz. Since the zero-intensity bunch length is $\tau_{4\sigma}\approx 0.75$ ns, we get $\omega_r \sigma \approx 1.7$. As was shown in the original work of Oide and Yokoya~\cite{OY:1990}, the results of self-consistent calculations are close to those obtained neglecting PWD and rf nonlinearity $\omega_r \sigma>0.4$.

In the recent publication~\cite{MetralMigliorati:2020}, the instability analysis was performed for the SPS parameters and the azimuthal mode-coupling instability was found assuming a single broad-band resonator impedance with the resonant frequency of 1 GHz. Additionally, a ten times smaller rf frequency compared to actual rf frequency was used to exclude the contribution of rf nonlinearity. The computed instability threshold was slightly higher than the one obtained in simulations. Our self-consistent analysis, instead, finds the radial mode-coupling instability. Since $\omega_r \sigma\approx4.3$, two types of semianalytical calculations give similar results as expected. Nevertheless, the growth rates as a function of intensity could be very different.

\subsection{Comparison with measurements}
\begin{figure}[tb]
\begin{center}
\includegraphics{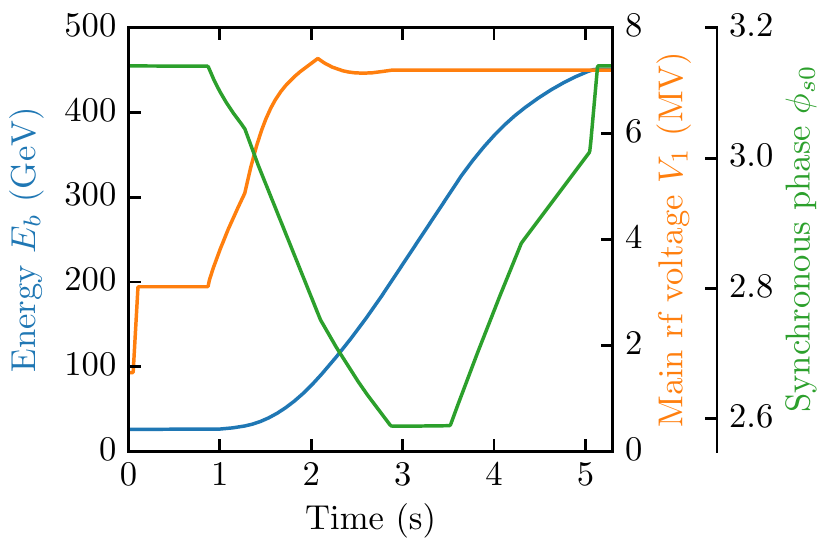}
\caption{The SPS acceleration cycle used during beam measurements~\cite{Lasheen:2017}: beam energy (blue), main rf voltage (orange), and synchronous phase (green).}
\label{fig:cycle}
\end{center}
\end{figure}

\begin{figure}[tb]
\begin{center}
\includegraphics{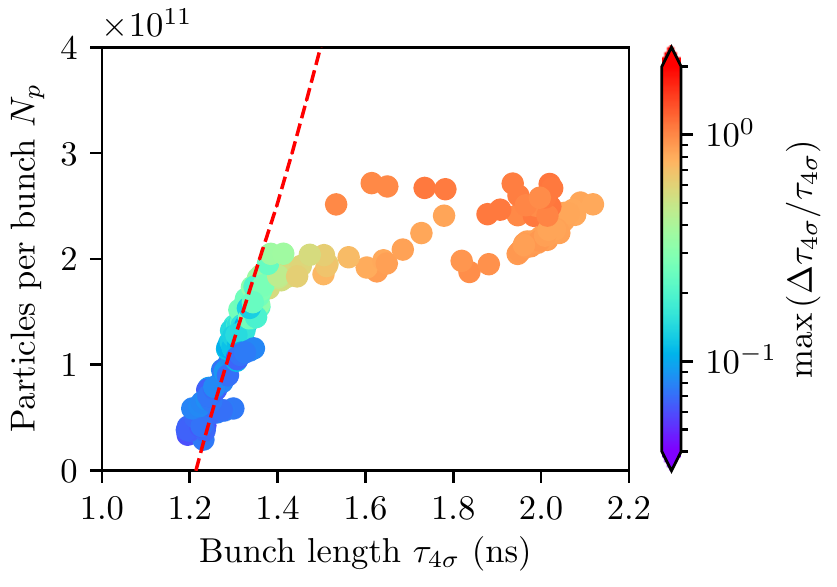}

\caption{
The maximum oscillation amplitude of the bunch length at the SPS flat top for different intensities and bunch lengths obtained from measurements~\cite{Lasheen:2017}.
The red dashed line corresponds to bunch parameters with a constant emittance of $\epsilon=0.33$~eVs.}
\label{fig:blond_vs_meas}
\end{center}
\end{figure}

During previous instability studies~\cite{Lasheen:2017}, macroparticle simulations were compared with dedicated measurements, where bunches with different intensities and approximately constant longitudinal emittance were injected into the SPS and accelerated following the cycle shown in Fig.~\ref{fig:cycle}. In general, the results of simulations through the ramp are very close to the observations if the latest impedance model is applied~\cite{Repond:2019}.
However, the impact of the transfer function of the measurement system on the bunch profiles was not yet taken into account.
In Fig.~\ref{fig:blond_vs_meas} the corrected data for bunch lengths and intensities is presented. 
Bunches with intensity below $1.2\times10^{11}$ remain stable for the whole cycle, which is consistent with predictions from simulations~(Fig.~\ref{fig:blond_only}) and seminalytical analysis~(Fig.~\ref{fig:melody_only}). There is a `mild' instability for the intensity range $1.2-1.8\times10^{11}$ already during the acceleration. Note that the bunch length as a function of intensity approximately follows the line for the total longitudinal emittance of 0.33~eVs up to $N_p\approx1.8\times10^{11}$ (red dashed line).
Bunches with even higher intensities suffer from a strong instability, which leads to the uncontrolled longitudinal emittance blowup. The presence of instability during the ramp is also consistent with simulations for the full acceleration cycle~\cite{Lasheen:2017,Repond:2019}. Still, stability maps computed for the SPS flat-top energy are not sufficient to fully explain the observations.

To understand the instability during the acceleration we apply the eigenvalue analysis using matrix equations~(\ref{eq:matrix}), which is implemented in the code {\footnotesize MELODY}.
The main advantage of this approach is that we can obtain a snapshot of beam stability at any time during the cycle. In simulations, however, this is difficult since the energy changes every turn during acceleration, and only a very fast instability can be properly observed during the limited time of the small change in beam energy.
Note that in the stability analysis one also has to include the space-charge contribution to the total impedance model. At the injection energy of 26 GeV, the computed Im$Z/k$ is about $-1 \; \Omega$~\cite{Lasheen:2017}, but it drops by an order of magnitude at 100 GeV and thus can be neglected at higher energies.

\begin{figure}[bt]
\begin{center}
\includegraphics{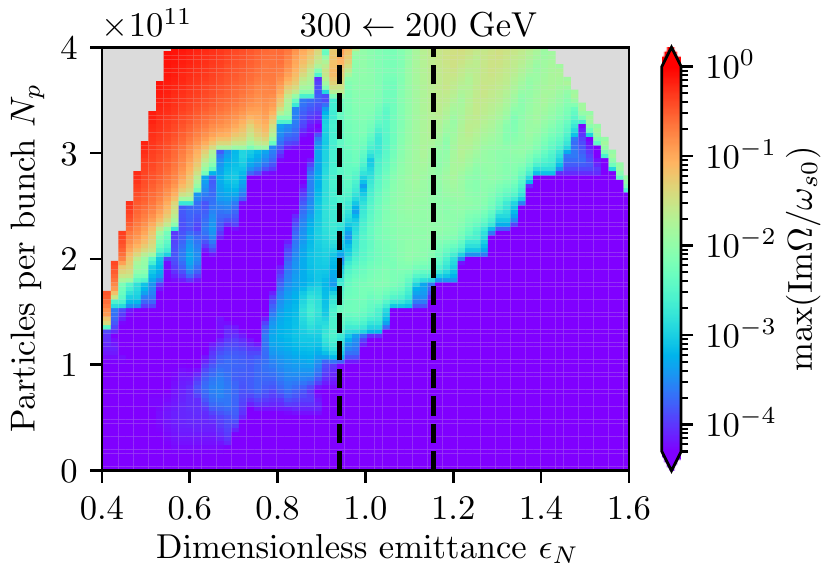}
\caption{Stability map applicable to various points of the acceleration (200 GeV to 300 GeV) obtained with {\footnotesize MELODY} for $\phi_{s0} = 2.56$. Background color describes the instability growth rates. Two vertical dashed lines indicate the range of the dimensionless emittances for the total longitudinal emittance $\epsilon = 0.33$~eVs.}
\label{fig:map_ramp}
\end{center}
\end{figure}

As was discussed in Sec.~\ref{subsec: Vlasov Equation}, the single-bunch stability for a given impedance model depends only on few dimensionless parameters. In the single-rf case these are $\zeta$, $\epsilon_N$, and $\phi_{s0}$.
Thanks to the systematic analysis performed by Gadioux~\cite{Gadioux:2020}, we have found that stability maps at intermediate energies through the cycle have similar nonmonotonic behavior than the one at the flat top with the unstable `island'.
In Fig.~\ref{fig:map_ramp}, as an example, the map of bunch intensity versus the dimensionless emittance obtained from calculations with code {\footnotesize MELODY}  is shown for $\phi_{s0} = 2.59$. It illustrates the beam stability during acceleration from 200 to 300~GeV since $\phi_{s0}$ and $V_1$ are constant  (see Fig.~\ref{fig:cycle}).
For longitudinal emittance of 0.33~eVs, the bunches with an intensity above $1.2\times 10^{11}$ will enter the unstable `island' where the radial mode-coupling instability occurs. Depending on the time spent in that region, a fast enough instability will be able to develop. For even higher intensities, the bunches will become unstable even earlier in the ramp. 
Thus, we can conclude that the instabilities observed in the dedicated measurements during the ramp are indeed caused by the coupling of the radial modes.

\section{Mitigation of the SPS instabilities}\label{sec: mitigation}
In this section, we discuss the ways applied to improve beam stability in the SPS.
We saw above that the type of single-bunch instability and its threshold strongly depend on the  synchrotron frequency distribution $\omega_s(\E)$. This function can be significantly modified by the PWD, determined in turn by the SPS impedance model, and a HH rf system. The two stabilization methods based on this effect are used in the SPS and considered below.
\subsection{Impact of the SPS impedance reduction}

\begin{figure}[tb]
\begin{center}
\includegraphics{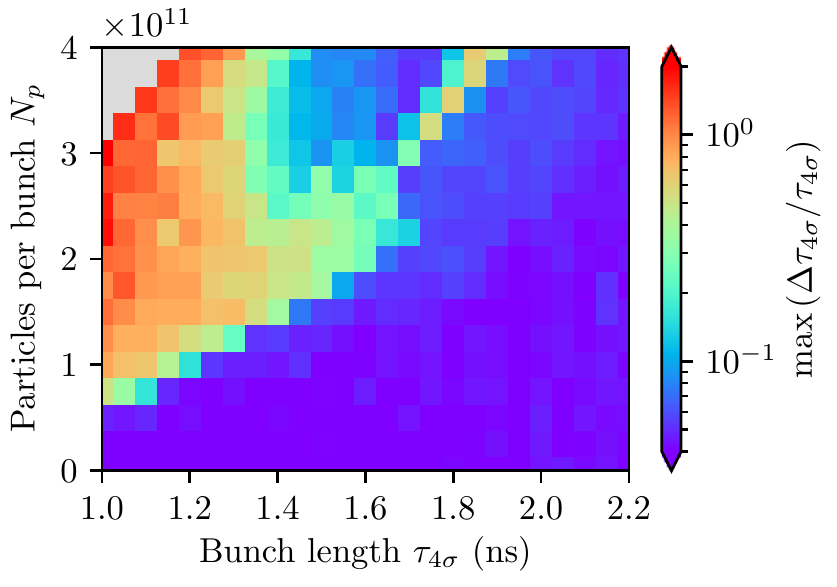}
\caption{Longitudinal stability map obtained from {\footnotesize BLonD} simulations for the SPS impedance model after LS2 (orange curve in Fig.~\ref{fig:sps_impedance}) and single-rf configuration. Background colors indicate the maximum relative oscillation amplitude of the bunch length. The SPS parameters are according to Table~\ref{tab: SPS_parameters}.}
\label{fig:blond_future}
\end{center}
\end{figure}

As an injector for the Large Hadron Collider (LHC), the SPS has been majorly upgraded during the last long shutdown (LS2) to be able to produce the multi-bunch beam required for the High Luminosity (HL)-LHC~\cite{LIU2014,Shaposhnikova2011,Shaposhnikova2016}. In particular, an impedance reduction campaign~\cite{Shaposhnikova2016} has been performed, involving shielding of the specific type of vacuum flanges,  200~MHz rf system upgrade, and damping of its higher-order mode around 630 MHz by a factor of three.
A significant increase of instability thresholds for single-bunch and  LHC-type beams in the double-rf system was expected from simulations and already observed for available intensities~\cite{ipac2022VK}. However, stability of a single bunch in a single-rf configuration could be degraded for some bunch parameters~\cite{Radvilas:2015}. Indeed, the top plot in Fig.~\ref{fig:blond_future}  demonstrates that the stable region on the left-hand side of the unstable `island' now disappeared. 

\begin{figure}[bt]
\begin{center}
\includegraphics{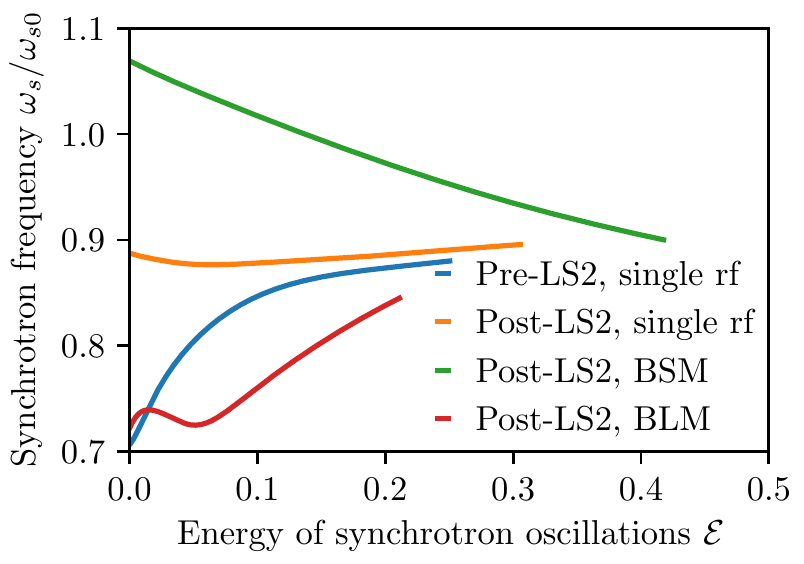}
\caption{Synchrotron frequency distributions for different impedance models and rf configurations at the intensity $N_p=2\times 10^{11}$ and the bunch length $\tau_{4\sigma}=1.3$ ns.}
\label{fig:fs_distr_future2021}
\end{center}
\end{figure}

This effect can be again understood from the new shapes of $\omega_s(\E)$ shown in Fig.~\ref{fig:fs_distr_future2021}.
Bunches with an intensity of~$2\times10^{11}$ and a length of~1.3~ns were stable before impedance reduction because $\omega_s(\E)$ was a monotonic function (blue). After LS2 an overlap of synchrotron frequencies for different $\E$ (orange) occurs, which results in the radial mode-coupling instability.

\subsection{High-harmonic rf system}

Double-rf operation in the SPS significantly increases the stability of all beams. The 800~MHz voltage is always applied in BSM, while there is no improvement of stability in BLM.
The latter is practically limited by the extremely accurate programming of the relative phase $\Phi_4$ in Eq.~(\ref{eq:drf}) between the 200~MHz and 800~MHz rf systems during the entire cycle~\cite{Shaposhnikova1998}. Additionally, a strong PWD, especially due to the resistive part of the SPS impedance, also introduces a shift of the synchronous phase $\Delta \phi_s$, such that the relative phase $\Phi_4$ computed without intensity effects is not optimal anymore. This leads to a nonmonotonic synchrotron frequency as a function of $\E$ (Fig.~\ref{fig:fs_distr_future2021}, red) and radial mode-coupling instability can occur. To suppress this instability for given bunch parameters, the relative phase $\Phi_4$ must be precisely tuned to reestablish $\omega_s(\E)$ as a monotonic function. If the intensity is very low, or the resistive impedance is negligible, the instability can be avoided, which is consistent with observations in another CERN synchrotron, the PS Booster (PSB)~\cite{Albright2021}. It is also known, that HH cavities operating in BLM allow raising the instability threshold of electron bunches (e.g.,~\cite{Mosnier:1999}). In particular, passive rf~cavities can be used with a purely beam-induced rf voltage. To obtain a flat bunch profile, the amplitude and phase of the rf voltage are controlled by tuning the cavity center-frequency. In this case, a weak radial mode-coupling instability due to nonmonotonic synchrotron frequency distribution is suppressed by the synchrotron radiation damping.

\begin{figure}[tb]
\begin{center}
\includegraphics{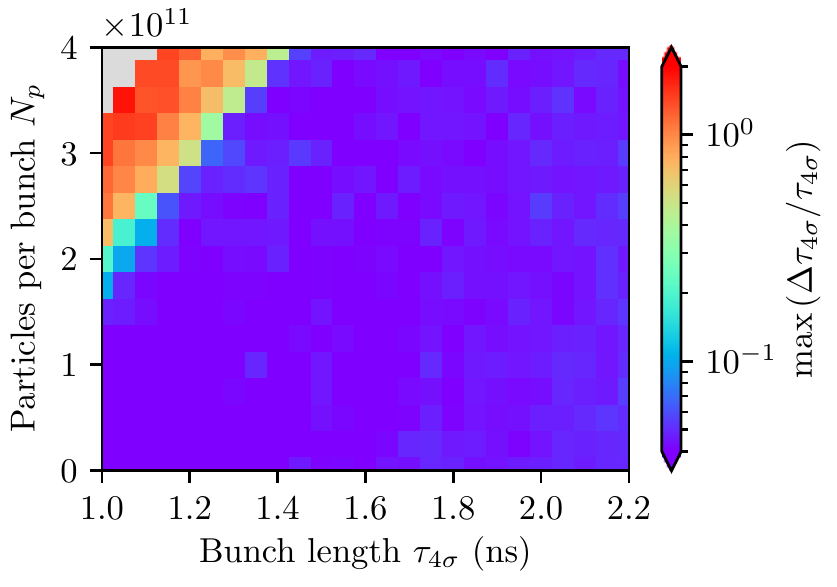}
\caption{Longitudinal stability map obtained from {\footnotesize BLonD} simulations for the SPS impedance model after LS2 (orange curve in Fig.~\ref{fig:sps_impedance}) and double-rf configuration (BSM with a voltage ratio $r=0.1$). Background colors indicate the maximum relative oscillation amplitude of the bunch length. The SPS parameters are from Table~\ref{tab: SPS_parameters}.}
\label{fig:blond_future_double}
\end{center}
\end{figure}

The operation in BSM is much less sensitive to the relative phase between rf systems and can help to suppress the harmful impact of PWD (Fig.~~\ref{fig:fs_distr_future2021}, green).
This rf configuration is routinely used in the SPS to deliver high-intensity bunches to the AWAKE experiment~\cite{Caldwell2016}. For example, for voltage ratio of $r =0.1$ a significant improvement in beam stability, in comparison to a single-rf case (Fig.~\ref{fig:blond_future_double}). Typically, the instability threshold for short bunches gets higher for a larger ratio of rf voltages $r$, however, special attention is necessary for long bunches. These bunches may have a nonmonotonic synchrotron distribution for $r>1/n$ and $n>2$ even at zero intensity. The instability mechanism in this configuration is under study.

\section{Conclusion}\label{sec: conclusions}

We have demonstrated the existence of two different mechanisms of longitudinal single-bunch instability in the SPS: radial and azimuthal mode-coupling. The  results of fully self-consistent semianalytic calculations done using the code {\footnotesize MELODY} are able to explain all  relevant observations in the SPS, including nonmonotonic dependence of thresholds on intensity and the presence of the unstable `island'. 
The theoretical predictions also agree well with the performed macroparticle simulations.

Previously, radial-mode coupling instability was obtained for very short, electron, bunches, when potential-well distortion plays an important role but rf nonlinearity can be neglected. 
We have shown that neglecting rf nonlinearity for long proton bunches leads to underestimation of the real instability threshold up to a factor of five.
With intensity increase, radial mode-coupling can appear simultaneously for several azimuthal modes and then has a signature of microwave instability. 

Another instability type, azimuthal mode-coupling, is also possible in the SPS, but for shorter bunches. In this case, the corresponding modes lose Landau damping (moving outside the incoherent frequency bands) and can become coupled above a certain beam intensity. This instability has a similar threshold but seems to be significantly weaker than the one obtained previously using non-self-consistent analysis (without PWD and rf nonlinearity). It can be suppressed by a small change in the particle distribution or even increase of intensity. Additionally, we also observed a `mixed' mode-coupling instability, which phase-space perturbation involves several azimuthal modes, and it has higher growth rates.

Longitudinal instabilities in the SPS are cured by operating in the bunch-shortening mode of the double-rf configuration, as it removes the nonmonotonic behavior of the synchrotron frequency as a function of synchrotron oscillation energy at the bunch center. Thus, for example, applying the 4th harmonic voltage at 10\% of the main-rf voltage  improves the stability of short bunches by almost a factor of three, especially after the recent impedance reduction campaign. Longer bunches should be treated with caution due to nonmonotonic behavior of synchrotron frequency distribution occuring with 4th harmonic rf system.

\begin{acknowledgments}
I wish to acknowledge Elena Shaposhnikova for fruitful discussions and valuable input to this manuscript. I am grateful to Heiko Damerau for his useful comments and continuous support. Special thanks also to Alexandre Lasheen, who provided the analyzed data from the dedicated experiments. Last but not least, I would like to thank Maxime Gadioux for his excellent work during the CERN Summer Student programme 2020.
\end{acknowledgments}


\appendix
\section{\label{annex} Azimuthal mode-coupling instability}
The azimuthal mode-coupling instability mechanism was proposed by Sacherer in~\cite{Sacherer:1977}. Here, we demonstrate that this is the only instability mechanism when rf nonlinearity and PWD are neglected.
In this case, the functions $I_{mk}$ according to Eq.~(\ref{eq: Imk}) can be evaluated analytically. They become \[I_{mk}(\E)=i^m J_m\left(k\sqrt{2\E}/h\right)\], where $J_m(x)$ is the Bessel function of the first kind and the order $m$. Then, the nominator of the integrand of the spectral harmonics $\Tilde{\lambda}_k(\Omega)$ from Eq.~(\ref{eq:line_density_pert_harm_OY}) becomes
\begin{align}
    \label{eq:line_density_pert_harm_azimuthal_mc}
    \sum_{m=1}^{\infty}C_m(\E,\Omega)& I^*_{mk}(\E) \nonumber \\
    = \sum_{p =1 }^{\infty}(-1)^p &\left\{ \text{Re}[C_{2p}(\E,\Omega)] J_{2p}\left(\frac{k}{h}\sqrt{2\E}\right) \nonumber \right.\\
    &\left. - \text{Im}[C_{2p-1}(\E,\Omega)] J_{2p-1}\left(\frac{k}{h}\sqrt{2\E}\right)\right\} \nonumber \\
    +i\sum_{p =1 }^{\infty}(-1)^p &\left\{ \text{Im}[C_{2p}(\E,\Omega)] J_{2p}\left(\frac{k}{h}\sqrt{2\E}\right) \nonumber \right.\\
    &\left. + \text{Re}[C_{2p-1}(\E,\Omega)] J_{2p-1}\left(\frac{k}{h}\sqrt{2\E}\right)\right\}.
\end{align}
Based on the Bessel function property for the integer argument $J_m(-x)=(-1)^m J_m(x)$, $\left|\Tilde{\lambda}_k(\Omega)\right|^2$ is a symmetric function of $k$ if either $\text{Re}\left[C_m(\E,\Omega)\right]=0$ or $\text{Im}\left[C_m(\E,\Omega)\right]=0$ for all azimuthal modes $m$. Thus, $\text{Im}\Omega=0$ according to Eq.~(\ref{eq:growth_rate}) and the mode is stable. The instability requires that for at least one pair of azimuthal modes $m$ and $m+1$ the eigenfunctions $\text{Re}\left[C_m(\E,\Omega)\right]\neq0$ and $\text{Im}\left[C_m(\E,\Omega)\right]\neq0$. Then, at least one element of the sum in the integrand is non-zero and $\left|\Tilde{\lambda}_k(\Omega)\right|^2$ is asymmetric function of $k$, which is necessary condition for the instability. If only single azimuthal mode $m$ has nonzero eigenfuctions $\text{Re}\left[C_m(\E,\Omega)\right]$ and $\text{Im}\left[C_m(\E,\Omega)\right]$, the symmetry of $\left|\Tilde{\lambda}_k(\Omega)\right|^2$ is preserved and bunch remains stable.

\nocite{apsrev42Control}
\bibliographystyle{apsrev4-2}
\bibliography{apssamp.bib}

\begin{thebibliography}{39}%
\makeatletter
\providecommand \@ifxundefined [1]{%
 \@ifx{#1\undefined}
}%
\providecommand \@ifnum [1]{%
 \ifnum #1\expandafter \@firstoftwo
 \else \expandafter \@secondoftwo
 \fi
}%
\providecommand \@ifx [1]{%
 \ifx #1\expandafter \@firstoftwo
 \else \expandafter \@secondoftwo
 \fi
}%
\providecommand \natexlab [1]{#1}%
\providecommand \enquote  [1]{``#1''}%
\providecommand \bibnamefont  [1]{#1}%
\providecommand \bibfnamefont [1]{#1}%
\providecommand \citenamefont [1]{#1}%
\providecommand \href@noop [0]{\@secondoftwo}%
\providecommand \href [0]{\begingroup \@sanitize@url \@href}%
\providecommand \@href[1]{\@@startlink{#1}\@@href}%
\providecommand \@@href[1]{\endgroup#1\@@endlink}%
\providecommand \@sanitize@url [0]{\catcode `\\12\catcode `\$12\catcode
  `\&12\catcode `\#12\catcode `\^12\catcode `\_12\catcode `\%12\relax}%
\providecommand \@@startlink[1]{}%
\providecommand \@@endlink[0]{}%
\providecommand \url  [0]{\begingroup\@sanitize@url \@url }%
\providecommand \@url [1]{\endgroup\@href {#1}{\urlprefix }}%
\providecommand \urlprefix  [0]{URL }%
\providecommand \Eprint [0]{\href }%
\providecommand \doibase [0]{https://doi.org/}%
\providecommand \selectlanguage [0]{\@gobble}%
\providecommand \bibinfo  [0]{\@secondoftwo}%
\providecommand \bibfield  [0]{\@secondoftwo}%
\providecommand \translation [1]{[#1]}%
\providecommand \BibitemOpen [0]{}%
\providecommand \bibitemStop [0]{}%
\providecommand \bibitemNoStop [0]{.\EOS\space}%
\providecommand \EOS [0]{\spacefactor3000\relax}%
\providecommand \BibitemShut  [1]{\csname bibitem#1\endcsname}%
\let\auto@bib@innerbib\@empty
\bibitem [{\citenamefont {Sacherer}(1977)}]{Sacherer:1977}%
  \BibitemOpen
  \bibfield  {author} {\bibinfo {author} {\bibfnamefont {F.~J.}\ \bibnamefont
  {Sacherer}},\ }\bibfield  {title} {\bibinfo {title} {{Bunch lengthening and
  microwave instability}},\ }\href {https://doi.org/10.1109/TNS.1977.4328955}
  {\bibfield  {journal} {\bibinfo  {journal} {IEEE Trans. Nucl. Sci.}\ }\textbf
  {\bibinfo {volume} {24}},\ \bibinfo {pages} {1393} (\bibinfo {year}
  {1977})}\BibitemShut {NoStop}%
\bibitem [{\citenamefont {{Besnier}}(1979)}]{Besnier:1979}%
  \BibitemOpen
  \bibfield  {author} {\bibinfo {author} {\bibfnamefont {G.}~\bibnamefont
  {{Besnier}}},\ }\bibfield  {title} {\bibinfo {title} {Stabilit\'e des
  oscillations longitudinales d'un faisceau groupe se propageant dans une
  chambre a vide d'impedance reactive},\ }\href
  {https://doi.org/10.1016/0029-554X(79)90241-6} {\bibfield  {journal}
  {\bibinfo  {journal} {Nucl. Instrum. Methods}\ }\textbf {\bibinfo {volume}
  {164}},\ \bibinfo {pages} {235} (\bibinfo {year} {1979})}\BibitemShut
  {NoStop}%
\bibitem [{\citenamefont {Wang}\ and\ \citenamefont
  {Pellegrini}(1980)}]{Wang1980}%
  \BibitemOpen
  \bibfield  {author} {\bibinfo {author} {\bibfnamefont {J.~M.}\ \bibnamefont
  {Wang}}\ and\ \bibinfo {author} {\bibfnamefont {C.}~\bibnamefont
  {Pellegrini}},\ }\bibinfo {title} {On the condition for a single bunch high
  frequency fast blow-up},\ in\ \href
  {https://doi.org/10.1007/978-3-0348-5540-2_72} {\emph {\bibinfo {booktitle}
  {11th International Conference on High-Energy Accelerators, Geneva,
  Switzerland}}},\ \bibinfo {editor} {edited by\ \bibinfo {editor}
  {\bibfnamefont {W.~S.}\ \bibnamefont {Newman}}}\ (\bibinfo  {publisher}
  {Birkh{\"a}user Basel},\ \bibinfo {address} {Basel, Switzerland},\ \bibinfo
  {year} {1980})\ pp.\ \bibinfo {pages} {554--561}\BibitemShut {NoStop}%
\bibitem [{\citenamefont {Chin}\ \emph {et~al.}(1983)\citenamefont {Chin},
  \citenamefont {Satoh},\ and\ \citenamefont {Yokoya}}]{YHChin:1983}%
  \BibitemOpen
  \bibfield  {author} {\bibinfo {author} {\bibfnamefont {Y.~H.}\ \bibnamefont
  {Chin}}, \bibinfo {author} {\bibfnamefont {K.}~\bibnamefont {Satoh}},\ and\
  \bibinfo {author} {\bibfnamefont {K.}~\bibnamefont {Yokoya}},\ }\bibfield
  {title} {\bibinfo {title} {{Instability of a bunched beam with synchrotron
  frequency spread}},\ }\href@noop {} {\bibfield  {journal} {\bibinfo
  {journal} {Part. Accel.}\ }\textbf {\bibinfo {volume} {13}},\ \bibinfo
  {pages} {45} (\bibinfo {year} {1983})}\BibitemShut {NoStop}%
\bibitem [{\citenamefont {Krinsky}\ and\ \citenamefont
  {Wang}(1983)}]{Krinsky1983}%
  \BibitemOpen
  \bibfield  {author} {\bibinfo {author} {\bibfnamefont {S.}~\bibnamefont
  {Krinsky}}\ and\ \bibinfo {author} {\bibfnamefont {J.~M.}\ \bibnamefont
  {Wang}},\ }\bibfield  {title} {\bibinfo {title} {Longitudinal instabilities
  with a non-harmonic rf potential},\ }\href
  {https://doi.org/10.1109/TNS.1983.4332858} {\bibfield  {journal} {\bibinfo
  {journal} {IEEE Trans. Nucl. Sci.}\ }\textbf {\bibinfo {volume} {30}},\
  \bibinfo {pages} {2492} (\bibinfo {year} {1983})}\BibitemShut {NoStop}%
\bibitem [{\citenamefont {Laclare}(1987)}]{Laclare:1987}%
  \BibitemOpen
  \bibfield  {author} {\bibinfo {author} {\bibfnamefont {J.~L.}\ \bibnamefont
  {Laclare}},\ }\href {https://doi.org/10.5170/CERN-1987-003-V-1.264} {\emph
  {\bibinfo {title} {{Bunched beam coherent instabilities}}}},\ \bibinfo {type}
  {Tech. Rep.}\ \bibinfo {number} {CERN-1987-003-V-1.264}\ (\bibinfo
  {institution} {CERN},\ \bibinfo {address} {Geneva, Switzerland},\ \bibinfo
  {year} {1987})\BibitemShut {NoStop}%
\bibitem [{\citenamefont {Garnier}(1987)}]{Garnier:1987}%
  \BibitemOpen
  \bibfield  {author} {\bibinfo {author} {\bibfnamefont {J.~P.}\ \bibnamefont
  {Garnier}},\ }\emph {\bibinfo {title} {Instabilit{\'e}s coh{\'e}rentes dans
  les acc{\'e}l{\'e}rateurs circulaires}},\ \href@noop {} {Ph.D. thesis},\
  \bibinfo  {school} {Institut National Polytechnique de Grenoble, Grenoble,
  France} (\bibinfo {year} {1987})\BibitemShut {NoStop}%
\bibitem [{\citenamefont {Oide}\ and\ \citenamefont {Yokoya}(1990)}]{OY:1990}%
  \BibitemOpen
  \bibfield  {author} {\bibinfo {author} {\bibfnamefont {K.}~\bibnamefont
  {Oide}}\ and\ \bibinfo {author} {\bibfnamefont {K.}~\bibnamefont {Yokoya}},\
  }\href@noop {} {\emph {\bibinfo {title} {{Longitudinal single bunch
  instability in electron storage rings}}}},\ \bibinfo {type} {Tech. Rep.}\
  \bibinfo {number} {KEK-Preprint-90-10}\ (\bibinfo  {institution} {KEK,
  Tsukuba, Japan},\ \bibinfo {year} {1990})\BibitemShut {NoStop}%
\bibitem [{\citenamefont {Oide}(1995)}]{Oide:1995}%
  \BibitemOpen
  \bibfield  {author} {\bibinfo {author} {\bibfnamefont {K.}~\bibnamefont
  {Oide}},\ }\bibfield  {title} {\bibinfo {title} {{A Mechanism of Longitudinal
  Single-Bunch Instability in Storage Rings}},\ }\href
  {https://cds.cern.ch/record/276015} {\bibfield  {journal} {\bibinfo
  {journal} {Part. Accel.}\ }\textbf {\bibinfo {volume} {51}},\ \bibinfo
  {pages} {43} (\bibinfo {year} {1995})}\BibitemShut {NoStop}%
\bibitem [{\citenamefont {Chao}\ \emph {et~al.}(1995)\citenamefont {Chao},
  \citenamefont {Chen},\ and\ \citenamefont {Oide}}]{Chao:1994}%
  \BibitemOpen
  \bibfield  {author} {\bibinfo {author} {\bibfnamefont {A.}~\bibnamefont
  {Chao}}, \bibinfo {author} {\bibfnamefont {Bo}~\bibnamefont {Chen}},\ and\
  \bibinfo {author} {\bibfnamefont {K.}~\bibnamefont {Oide}},\ }\bibfield
  {title} {\bibinfo {title} {{A weak microwave instability with potential well
  distortion and radial mode coupling}},\ }in\ \href
  {https://doi.org/10.1109/PAC.1995.505777} {\emph {\bibinfo {booktitle}
  {Proceedings of 16th Particle Accelerator Conference, Dallas, Texas, 1995}}}\
  (\bibinfo  {publisher} {IEEE, New York},\ \bibinfo {year} {1995})\ pp.\
  \bibinfo {pages} {3040--3042}\BibitemShut {NoStop}%
\bibitem [{\citenamefont {Ng}(1995)}]{Ng:1995}%
  \BibitemOpen
  \bibfield  {author} {\bibinfo {author} {\bibfnamefont {K.~Y.}\ \bibnamefont
  {Ng}},\ }\bibfield  {title} {\bibinfo {title} {Mode-coupling instability and
  bunch lengthening in proton machines},\ }in\ \href
  {https://doi.org/10.1109/PAC.1995.505756} {\emph {\bibinfo {booktitle}
  {Proceedings of 16th Particle Accelerator Conference, Dallas, Texas, 1995}}}\
  (\bibinfo  {publisher} {IEEE, New York},\ \bibinfo {year} {1995})\ pp.\
  \bibinfo {pages} {2977--2979}\BibitemShut {NoStop}%
\bibitem [{\citenamefont {{D'yachkov}}\ and\ \citenamefont
  {{Baartman}}(1995)}]{Dyachkov:1995}%
  \BibitemOpen
  \bibfield  {author} {\bibinfo {author} {\bibfnamefont {M.}~\bibnamefont
  {{D'yachkov}}}\ and\ \bibinfo {author} {\bibfnamefont {R.}~\bibnamefont
  {{Baartman}}},\ }\bibfield  {title} {\bibinfo {title} {{Longitudinal single
  bunch stability}},\ }\href@noop {} {\bibfield  {journal} {\bibinfo  {journal}
  {Part. Accel.}\ }\textbf {\bibinfo {volume} {50}},\ \bibinfo {pages} {105}
  (\bibinfo {year} {1995})}\BibitemShut {NoStop}%
\bibitem [{\citenamefont {Mosnier}(1999)}]{Mosnier:1999}%
  \BibitemOpen
  \bibfield  {author} {\bibinfo {author} {\bibfnamefont {A.}~\bibnamefont
  {Mosnier}},\ }\bibfield  {title} {\bibinfo {title} {Microwave instability in
  electron storage rings},\ }\href
  {https://doi.org/https://doi.org/10.1016/S0168-9002(99)00832-3} {\bibfield
  {journal} {\bibinfo  {journal} {Nucl. Instrum. Methods}\ }\textbf {\bibinfo
  {volume} {438}},\ \bibinfo {pages} {225} (\bibinfo {year}
  {1999})}\BibitemShut {NoStop}%
\bibitem [{\citenamefont {Cai}(2011)}]{Cai2011}%
  \BibitemOpen
  \bibfield  {author} {\bibinfo {author} {\bibfnamefont {Y.}~\bibnamefont
  {Cai}},\ }\bibfield  {title} {\bibinfo {title} {Linear theory of microwave
  instability in electron storage rings},\ }\href
  {https://doi.org/10.1103/PhysRevSTAB.14.061002} {\bibfield  {journal}
  {\bibinfo  {journal} {Phys. Rev. ST Accel. Beams}\ }\textbf {\bibinfo
  {volume} {14}},\ \bibinfo {pages} {061002} (\bibinfo {year}
  {2011})}\BibitemShut {NoStop}%
\bibitem [{\citenamefont {Lindberg}(2019)}]{Lindberg2017}%
  \BibitemOpen
  \bibfield  {author} {\bibinfo {author} {\bibfnamefont {R.~R.}\ \bibnamefont
  {Lindberg}},\ }\bibinfo {title} {Practical theory to compute the microwave
  instability threshold},\ in\ \href
  {https://doi.org/10.1142/9789813279612_0012} {\emph {\bibinfo {booktitle}
  {Nonlinear Dynamics and Collective Effects in Particle Beam Physics}}}\
  (\bibinfo {year} {2019})\ pp.\ \bibinfo {pages} {138--146},\ \bibinfo {note}
  {and \emph{in Proceedings of NOCE 2017 Workshop, Arcidosso, Italy,
  2017}}\BibitemShut {NoStop}%
\bibitem [{\citenamefont {Blednykh}\ \emph {et~al.}(2018)\citenamefont
  {Blednykh}, \citenamefont {Bacha}, \citenamefont {Bassi}, \citenamefont
  {Cheng}, \citenamefont {Chubar}, \citenamefont {Derbenev}, \citenamefont
  {Lindberg}, \citenamefont {Rakitin}, \citenamefont {Smaluk}, \citenamefont
  {Zhernenkov}, \citenamefont {Chen-Wiegart},\ and\ \citenamefont
  {Wiegart}}]{Blednykh2018}%
  \BibitemOpen
  \bibfield  {author} {\bibinfo {author} {\bibfnamefont {A.}~\bibnamefont
  {Blednykh}}, \bibinfo {author} {\bibfnamefont {B.}~\bibnamefont {Bacha}},
  \bibinfo {author} {\bibfnamefont {G.}~\bibnamefont {Bassi}}, \bibinfo
  {author} {\bibfnamefont {W.}~\bibnamefont {Cheng}}, \bibinfo {author}
  {\bibfnamefont {O.}~\bibnamefont {Chubar}}, \bibinfo {author} {\bibfnamefont
  {A.}~\bibnamefont {Derbenev}}, \bibinfo {author} {\bibfnamefont
  {R.}~\bibnamefont {Lindberg}}, \bibinfo {author} {\bibfnamefont
  {M.}~\bibnamefont {Rakitin}}, \bibinfo {author} {\bibfnamefont
  {V.}~\bibnamefont {Smaluk}}, \bibinfo {author} {\bibfnamefont
  {M.}~\bibnamefont {Zhernenkov}}, \bibinfo {author} {\bibfnamefont {Y.~K.}\
  \bibnamefont {Chen-Wiegart}},\ and\ \bibinfo {author} {\bibfnamefont
  {L.}~\bibnamefont {Wiegart}},\ }\bibfield  {title} {\bibinfo {title} {New
  aspects of longitudinal instabilities in electron storage rings},\ }\href
  {https://doi.org/10.1038/s41598-018-30306-y} {\bibfield  {journal} {\bibinfo
  {journal} {Scientific Reports}\ }\textbf {\bibinfo {volume} {8}},\ \bibinfo
  {pages} {11918} (\bibinfo {year} {2018})}\BibitemShut {NoStop}%
\bibitem [{\citenamefont {M\'etral}\ and\ \citenamefont
  {Migliorati}(2020)}]{MetralMigliorati:2020}%
  \BibitemOpen
  \bibfield  {author} {\bibinfo {author} {\bibfnamefont {E.}~\bibnamefont
  {M\'etral}}\ and\ \bibinfo {author} {\bibfnamefont {M.}~\bibnamefont
  {Migliorati}},\ }\bibfield  {title} {\bibinfo {title} {Longitudinal and
  transverse mode coupling instability: Vlasov solvers and tracking codes},\
  }\href {https://doi.org/10.1103/PhysRevAccelBeams.23.071001} {\bibfield
  {journal} {\bibinfo  {journal} {Phys. Rev. Accel. Beams}\ }\textbf {\bibinfo
  {volume} {23}},\ \bibinfo {pages} {071001} (\bibinfo {year}
  {2020})}\BibitemShut {NoStop}%
\bibitem [{\citenamefont {Lasheen}(2017)}]{Lasheen:2017}%
  \BibitemOpen
  \bibfield  {author} {\bibinfo {author} {\bibfnamefont {A.~S.}\ \bibnamefont
  {Lasheen}},\ }\emph {\bibinfo {title} {Beam Measurements of the Longitudinal
  Impedance of the CERN Super Proton Synchrotron}},\ \href@noop {} {Ph.D.
  thesis},\ \bibinfo  {school} {Université Paris Saclay, Paris} (\bibinfo
  {year} {2017})\BibitemShut {NoStop}%
\bibitem [{SPS()}]{SPS:2021}%
  \BibitemOpen
  \href@noop {} {}\bibinfo {note} {CERN SPS Longitudinal Impedance Model,
  \url{https://gitlab.cern.ch/longitudinal-impedance/SPS}}\BibitemShut
  {NoStop}%
\bibitem [{Blo()}]{Blond:2021}%
  \BibitemOpen
  \href@noop {} {}\bibinfo {note} {{CERN Beam Longitudinal Dynamics code BLonD,
  \url{http://blond.web.cern.ch}}}\BibitemShut {NoStop}%
\bibitem [{\citenamefont {Repond}(2019)}]{Repond:2019}%
  \BibitemOpen
  \bibfield  {author} {\bibinfo {author} {\bibfnamefont {J.}~\bibnamefont
  {Repond}},\ }\emph {\bibinfo {title} {{Possible Mitigations of Longitudinal
  Intensity Limitations for HL-LHC Beam in the CERN SPS}}},\ \href
  {https://cds.cern.ch/record/2695204} {Ph.D. thesis},\ \bibinfo  {school}
  {Ecole Politechnique, Lausanne} (\bibinfo {year} {2019})\BibitemShut
  {NoStop}%
\bibitem [{\citenamefont {Karpov}()}]{Melody:2021}%
  \BibitemOpen
  \bibfield  {author} {\bibinfo {author} {\bibfnamefont {I.}~\bibnamefont
  {Karpov}},\ }\href@noop {} {}\bibinfo {note} {Matrix Equations for
  LOngitudinal beam DYnamics (MELODY) code,
  \url{https://gitlab.cern.ch/ikarpov/melody}}\BibitemShut {NoStop}%
\bibitem [{\citenamefont {{Burov}}()}]{burov2012van}%
  \BibitemOpen
  \bibfield  {author} {\bibinfo {author} {\bibfnamefont {A.}~\bibnamefont
  {{Burov}}},\ }\bibfield  {title} {\bibinfo {title} {{Van Kampen modes for
  bunch longitudinal motion}},\ }in\ \href@noop {} {\emph {\bibinfo {booktitle}
  {{Proceedings of 46th ICFA Advanced Beam Dynamics Workshop on High-Intensity
  and High-Brightness Hadron Beams, Morschach, Switzerland, 2010}}}},\ \Eprint
  {https://arxiv.org/abs/1207.5826} {arXiv:1207.5826 [physics.acc-ph]}
  \BibitemShut {NoStop}%
\bibitem [{\citenamefont {Chao}(1993)}]{AChao1993}%
  \BibitemOpen
  \bibfield  {author} {\bibinfo {author} {\bibfnamefont {A.W.}\ \bibnamefont
  {Chao}},\ }\bibinfo {title} {{Physics of Collective Beam Instabilities in
  High Energy Accelerators}}\ (\bibinfo  {publisher} {Wiley, New York},\
  \bibinfo {year} {1993})\ Chap.~\bibinfo {chapter} {6}, pp.\ \bibinfo {pages}
  {273--278}\BibitemShut {NoStop}%
\bibitem [{\citenamefont {Karpov}\ \emph {et~al.}(2021)\citenamefont {Karpov},
  \citenamefont {Argyropoulos},\ and\ \citenamefont
  {Shaposhnikova}}]{Karpov:2021}%
  \BibitemOpen
  \bibfield  {author} {\bibinfo {author} {\bibfnamefont {I.}~\bibnamefont
  {Karpov}}, \bibinfo {author} {\bibfnamefont {T.}~\bibnamefont
  {Argyropoulos}},\ and\ \bibinfo {author} {\bibfnamefont {E.}~\bibnamefont
  {Shaposhnikova}},\ }\bibfield  {title} {\bibinfo {title} {Thresholds for loss
  of landau damping in longitudinal plane},\ }\href
  {https://doi.org/10.1103/PhysRevAccelBeams.24.011002} {\bibfield  {journal}
  {\bibinfo  {journal} {Phys. Rev. Accel. Beams}\ }\textbf {\bibinfo {volume}
  {24}},\ \bibinfo {pages} {011002} (\bibinfo {year} {2021})}\BibitemShut
  {NoStop}%
\bibitem [{\citenamefont {{Van Kampen}}(1955)}]{vKampen1}%
  \BibitemOpen
  \bibfield  {author} {\bibinfo {author} {\bibfnamefont {N.~G.}\ \bibnamefont
  {{Van Kampen}}},\ }\bibfield  {title} {\bibinfo {title} {On the theory of
  stationary waves in plasmas},\ }\href
  {https://doi.org/https://doi.org/10.1016/S0031-8914(55)93068-8} {\bibfield
  {journal} {\bibinfo  {journal} {Physica (Utrecht)}\ }\textbf {\bibinfo
  {volume} {21}},\ \bibinfo {pages} {949--963} (\bibinfo {year}
  {1955})}\BibitemShut {NoStop}%
\bibitem [{\citenamefont {{Van Kampen}}(1957)}]{vKampen2}%
  \BibitemOpen
  \bibfield  {author} {\bibinfo {author} {\bibfnamefont {N.~G.}\ \bibnamefont
  {{Van Kampen}}},\ }\bibfield  {title} {\bibinfo {title} {The dispersion
  equation for plasma waves},\ }\href
  {https://doi.org/https://doi.org/10.1016/S0031-8914(57)93718-7} {\bibfield
  {journal} {\bibinfo  {journal} {Physica (Utrecht)}\ }\textbf {\bibinfo
  {volume} {23}},\ \bibinfo {pages} {641--650} (\bibinfo {year}
  {1957})}\BibitemShut {NoStop}%
\bibitem [{\citenamefont {Case}(1959)}]{Case1959}%
  \BibitemOpen
  \bibfield  {author} {\bibinfo {author} {\bibfnamefont {K.~M.}\ \bibnamefont
  {Case}},\ }\bibfield  {title} {\bibinfo {title} {Plasma oscillations},\
  }\href {https://doi.org/https://doi.org/10.1016/0003-4916(59)90029-6}
  {\bibfield  {journal} {\bibinfo  {journal} {Annals of Physics}\ }\textbf
  {\bibinfo {volume} {7}},\ \bibinfo {pages} {349} (\bibinfo {year}
  {1959})}\BibitemShut {NoStop}%
\bibitem [{\citenamefont {Radvilas}(2015)}]{Radvilas:2015}%
  \BibitemOpen
  \bibfield  {author} {\bibinfo {author} {\bibfnamefont {E.}~\bibnamefont
  {Radvilas}},\ }\href {https://cds.cern.ch/record/2044554} {\emph {\bibinfo
  {title} {{Simulations of Single-Bunch Instability on Flat Top}}}},\ \bibinfo
  {type} {Tech. Rep.}\ \bibinfo {number} {CERN-STUDENTS-Note-2015-048}\
  (\bibinfo  {institution} {CERN, Geneva, Switzerland},\ \bibinfo {year}
  {2015})\BibitemShut {NoStop}%
\bibitem [{\citenamefont {Benedikt}\ \emph {et~al.}(2004)\citenamefont
  {Benedikt}, \citenamefont {Collier}, \citenamefont {Mertens}, \citenamefont
  {Poole},\ and\ \citenamefont {Schindl}}]{LHCDR3}%
  \BibitemOpen
  \bibfield  {author} {\bibinfo {author} {\bibfnamefont {M.}~\bibnamefont
  {Benedikt}}, \bibinfo {author} {\bibfnamefont {P.}~\bibnamefont {Collier}},
  \bibinfo {author} {\bibfnamefont {V.}~\bibnamefont {Mertens}}, \bibinfo
  {author} {\bibfnamefont {J.}~\bibnamefont {Poole}},\ and\ \bibinfo {author}
  {\bibfnamefont {K.}~\bibnamefont {Schindl}},\ }\href
  {https://doi.org/10.5170/CERN-2004-003-V-3} {\emph {\bibinfo {title} {{LHC
  Design Report vol.3: The LHC Injector Chain}}}},\ CERN Yellow Reports:
  Monographs\ (\bibinfo  {publisher} {CERN},\ \bibinfo {address} {Geneva,
  Switzerland},\ \bibinfo {year} {2004})\BibitemShut {NoStop}%
\bibitem [{\citenamefont {Karpov}\ and\ \citenamefont
  {Gadioux}(2021)}]{KarpovGadioux2021}%
  \BibitemOpen
  \bibfield  {author} {\bibinfo {author} {\bibfnamefont {I.}~\bibnamefont
  {Karpov}}\ and\ \bibinfo {author} {\bibfnamefont {M.}~\bibnamefont
  {Gadioux}},\ }\bibfield  {title} {\bibinfo {title} {{Mechanism of
  Longitudinal Single-Bunch Instability in the CERN SPS}},\ }in\ \href
  {https://doi.org/10.18429/JACoW-IPAC2021-WEPAB227} {\emph {\bibinfo
  {booktitle} {Proceedings of 12th International Particle Accelerator
  Conference, Campinas, SP, Brazil, 2021}}}\ (\bibinfo  {publisher} {JACoW,
  Geneva, Switzerland},\ \bibinfo {year} {2021})\ pp.\ \bibinfo {pages}
  {3161--3164}\BibitemShut {NoStop}%
\bibitem [{\citenamefont {Gadioux}(2020)}]{Gadioux:2020}%
  \BibitemOpen
  \bibfield  {author} {\bibinfo {author} {\bibfnamefont {M.}~\bibnamefont
  {Gadioux}},\ }\href {https://cds.cern.ch/record/2742420} {\emph {\bibinfo
  {title} {{Evaluation of Longitudinal Single-Bunch Stability in the SPS and
  Bunch Optimisation for AWAKE}}}},\ \bibinfo {type} {Tech. Rep.}\ \bibinfo
  {number} {{CERN-STUDENTS-Note-2020-030}}\ (\bibinfo  {institution} {CERN,
  Geneva, Switzerland},\ \bibinfo {year} {2020})\BibitemShut {NoStop}%
\bibitem [{LIU()}]{LIU2014}%
  \BibitemOpen
  \href@noop {} {}\bibinfo {note} {LHC Injectors Upgrade, Technical Design
  Report, Vol. I: Protons, edited by J. Coupard, H. Damerau, A. Funken, R.
  Garoby, S. Gilardoni, B. Goddard, K. Hanke, A. Lombardi, D. Manglunki, M.
  Meddahi, B. Mikulec, G. Rumolo, E. Shaposhnikova, M. Vretenar
  CERN-ACC-2014-0337 (CERN, Geneva, 2014)}\BibitemShut {NoStop}%
\bibitem [{\citenamefont {Shaposhnikova}\ \emph {et~al.}(2011)\citenamefont
  {Shaposhnikova}, \citenamefont {Ciapala},\ and\ \citenamefont
  {Montesinos}}]{Shaposhnikova2011}%
  \BibitemOpen
  \bibfield  {author} {\bibinfo {author} {\bibfnamefont {E.}~\bibnamefont
  {Shaposhnikova}}, \bibinfo {author} {\bibfnamefont {E.}~\bibnamefont
  {Ciapala}},\ and\ \bibinfo {author} {\bibfnamefont {E.}~\bibnamefont
  {Montesinos}},\ }\bibfield  {title} {\bibinfo {title} {{Upgrade of the 200
  MHz rf system in the CERN SPS}},\ }in\ \href@noop {} {\emph {\bibinfo
  {booktitle} {Proceedings of 2nd International Particle Accelerator
  Conference, San Sebastian, Spain, 2011}}}\ (\bibinfo  {publisher} {JACoW,
  Geneva, Switzerland},\ \bibinfo {year} {2011})\BibitemShut {NoStop}%
\bibitem [{\citenamefont {Shaposhnikova}\ \emph {et~al.}(2016)\citenamefont
  {Shaposhnikova}, \citenamefont {Argyropoulos}, \citenamefont {Bohl},
  \citenamefont {Cruikshank}, \citenamefont {Goddard}, \citenamefont
  {Kaltenbacher}, \citenamefont {Lasheen}, \citenamefont {Perez~Espinos},
  \citenamefont {Repond}, \citenamefont {Salvant},\ and\ \citenamefont
  {Vollinger}}]{Shaposhnikova2016}%
  \BibitemOpen
  \bibfield  {author} {\bibinfo {author} {\bibfnamefont {E.}~\bibnamefont
  {Shaposhnikova}}, \bibinfo {author} {\bibfnamefont {T.}~\bibnamefont
  {Argyropoulos}}, \bibinfo {author} {\bibfnamefont {T.}~\bibnamefont {Bohl}},
  \bibinfo {author} {\bibfnamefont {P.}~\bibnamefont {Cruikshank}}, \bibinfo
  {author} {\bibfnamefont {B.}~\bibnamefont {Goddard}}, \bibinfo {author}
  {\bibfnamefont {T.}~\bibnamefont {Kaltenbacher}}, \bibinfo {author}
  {\bibfnamefont {A.}~\bibnamefont {Lasheen}}, \bibinfo {author} {\bibfnamefont
  {J.}~\bibnamefont {Perez~Espinos}}, \bibinfo {author} {\bibfnamefont
  {J.}~\bibnamefont {Repond}}, \bibinfo {author} {\bibfnamefont
  {B.}~\bibnamefont {Salvant}},\ and\ \bibinfo {author} {\bibfnamefont
  {C.}~\bibnamefont {Vollinger}},\ }\bibfield  {title} {\bibinfo {title}
  {{Removing Known SPS Intensity Limitations for High Luminosity LHC Goals}},\
  }in\ \href@noop {} {\emph {\bibinfo {booktitle} {Proceedings of 7th
  International Particle Accelerator Conference, Busan, Korea, 2016}}}\
  (\bibinfo  {publisher} {JACoW, Geneva, Switzerland},\ \bibinfo {year}
  {2016})\BibitemShut {NoStop}%
\bibitem [{\citenamefont {Kain}\ \emph {et~al.}(2022)\citenamefont {Kain} \emph
  {et~al.}}]{ipac2022VK}%
  \BibitemOpen
  \bibfield  {author} {\bibinfo {author} {\bibfnamefont {V.}~\bibnamefont
  {Kain}} \emph {et~al.},\ }\bibfield  {title} {\bibinfo {title} {{Achievements
  and Performance Prospects of the Upgraded LHC Injectors}},\ }in\ \href
  {https://doi.org/10.18429/JACoW-IPAC2022-WEIYGD1} {\emph {\bibinfo
  {booktitle} {Proceedings of 13th International Particle Accelerator
  Conference, Bangkok, Thailand, 2022}}}\ (\bibinfo  {publisher} {JACoW
  Publishing, Geneva, Switzerland},\ \bibinfo {year} {2022})\ pp.\ \bibinfo
  {pages} {1610--1615}\BibitemShut {NoStop}%
\bibitem [{\citenamefont {Bohl}\ \emph {et~al.}(1998)\citenamefont {Bohl},
  \citenamefont {Linnecar}, \citenamefont {Shaposhnikova},\ and\ \citenamefont
  {T\"uckmantel}}]{Shaposhnikova1998}%
  \BibitemOpen
  \bibfield  {author} {\bibinfo {author} {\bibfnamefont {T.}~\bibnamefont
  {Bohl}}, \bibinfo {author} {\bibfnamefont {T.~P.~R}\ \bibnamefont
  {Linnecar}}, \bibinfo {author} {\bibfnamefont {E.}~\bibnamefont
  {Shaposhnikova}},\ and\ \bibinfo {author} {\bibfnamefont {J.}~\bibnamefont
  {T\"uckmantel}},\ }\href {https://cds.cern.ch/record/360139} {\emph {\bibinfo
  {title} {{Study of different operating modes of the 4th rf harmonic Landau
  damping system in the CERN SPS}}}},\ \bibinfo {type} {Tech. Rep.}\ \bibinfo
  {number} {CERN-SL-98-026-RF}\ (\bibinfo  {institution} {CERN, Geneva,
  Switzerland},\ \bibinfo {year} {1998})\BibitemShut {NoStop}%
\bibitem [{\citenamefont {Albright}\ \emph {et~al.}(2021)\citenamefont
  {Albright}, \citenamefont {Antoniou}, \citenamefont {Asvesta}, \citenamefont
  {Bartosik}, \citenamefont {Bracco},\ and\ \citenamefont
  {Renner}}]{Albright2021}%
  \BibitemOpen
  \bibfield  {author} {\bibinfo {author} {\bibfnamefont {S.~C.~P.}\
  \bibnamefont {Albright}}, \bibinfo {author} {\bibfnamefont {F.}~\bibnamefont
  {Antoniou}}, \bibinfo {author} {\bibfnamefont {F.}~\bibnamefont {Asvesta}},
  \bibinfo {author} {\bibfnamefont {H.}~\bibnamefont {Bartosik}}, \bibinfo
  {author} {\bibfnamefont {C.}~\bibnamefont {Bracco}},\ and\ \bibinfo {author}
  {\bibfnamefont {E.}~\bibnamefont {Renner}},\ }\bibfield  {title} {\bibinfo
  {title} {{New Longitudinal Beam Production Methods in the CERN Proton
  Synchrotron Booster}},\ }in\ \href
  {https://doi.org/10.18429/JACoW-IPAC2021-THPAB183} {\emph {\bibinfo
  {booktitle} {Proceedings of 12th International Particle Accelerator
  Conference, Campinas, SP, Brazil, 2021}}}\ (\bibinfo  {publisher} {JACoW,
  Geneva, Switzerland},\ \bibinfo {year} {2021})\ pp.\ \bibinfo {pages}
  {4130--4133}\BibitemShut {NoStop}%
\bibitem [{\citenamefont {Caldwell}\ \emph {et~al.}(2016)\citenamefont
  {Caldwell}, \citenamefont {Adli}, \citenamefont {Amorim}, \citenamefont
  {Apsimon}, \citenamefont {Argyropoulos}, \citenamefont {Assmann},
  \citenamefont {Bachmann}, \citenamefont {Batsch}, \citenamefont {Bauche},
  \citenamefont {Olsen} \emph {et~al.}}]{Caldwell2016}%
  \BibitemOpen
  \bibfield  {author} {\bibinfo {author} {\bibfnamefont {A.}~\bibnamefont
  {Caldwell}}, \bibinfo {author} {\bibfnamefont {E.}~\bibnamefont {Adli}},
  \bibinfo {author} {\bibfnamefont {L.}~\bibnamefont {Amorim}}, \bibinfo
  {author} {\bibfnamefont {R.}~\bibnamefont {Apsimon}}, \bibinfo {author}
  {\bibfnamefont {T.}~\bibnamefont {Argyropoulos}}, \bibinfo {author}
  {\bibfnamefont {R.}~\bibnamefont {Assmann}}, \bibinfo {author} {\bibfnamefont
  {A.-M.}\ \bibnamefont {Bachmann}}, \bibinfo {author} {\bibfnamefont
  {F.}~\bibnamefont {Batsch}}, \bibinfo {author} {\bibfnamefont
  {J.}~\bibnamefont {Bauche}}, \bibinfo {author} {\bibfnamefont {V.K.~Berglyd}\
  \bibnamefont {Olsen}}, \emph {et~al.},\ }\bibfield  {title} {\bibinfo {title}
  {{Path to {AWAKE}: Evolution of the Concept}},\ }\href
  {https://doi.org/https://doi.org/10.1016/j.nima.2015.12.050} {\bibfield
  {journal} {\bibinfo  {journal} {Nucl. Instrum. Methods}\ }\textbf {\bibinfo
  {volume} {829}},\ \bibinfo {pages} {3} (\bibinfo {year} {2016})}\BibitemShut
  {NoStop}%
\end{thebibliography}%

\end{document}